ICRC 2013

# Pierre Auger Observatory and Telescope Array: Joint Contributions to the 33rd International Cosmic Ray Conference (ICRC 2013)

## The Telescope Array Collaboration

T. Abu-Zayyad[201], M. Allen[201], R. Anderson[201], R. Azuma[202], E. Barcikowski[201], J. W Belz[201], D. R. Bergman[201], S. A. Blake[201], R. Cady[201], M. J. Chae[203], B. G. Cheon[204], J. Chiba[205], M. Chikawa[206], W. R. Cho[207], T. Fujii[208], M. Fukushima[208,209], K. Goto[210], W. Hanlon[201], Y. Hayashi[210], N. Hayashida[211], K. Hibino[211], K. Honda[212], D. Ikeda[208], N. Inoue[213], T. Ishii[212], R. Ishimori[202], H. Ito[214], D. Ivanov[201], C. C. H. Jui[201], K. Kadota[216], F. Kakimoto[202], O. Kalashev[217], K. Kasahara[218], H. Kawai[219], S. Kawakami[210], S. Kawana[213], K. Kawata[208], E. Kido[208], H. B. Kim[204], J. H. Kim[201], J. H. Kim[204], S. Kitamura[202], Y. Kitamura[202], V. Kuzmin[217], Y. J. Kwon[207], J. Lan[201], J. P. Lundquist[201], K. Machida[212], K. Martens[209], T. Matsuda[220], T. Matsuyama[210], J. N. Matthews[201], M. Minamino[210], K. Mukai[212], I. Myers[201], K. Nagasawa[213], S. Nagataki[214], T. Nakamura[221], H. Nanpei[210], T. Nonaka[208], A. Nozato[206], S. Ogio[210], S. Oh[203], M. Ohnishi[208], H. Ohoka[208], K. Oki[208], T. Okuda[222], M. Ono[214], A. Oshima[210], S. Ozawa[218], I. H. Park[223], M. S. Pshirkov[224], D. C. Rodriguez[201], G. Rubtsov[217], D. Ryu[225], H. Sagawa[208], N. Sakurai[210], A. L. Sampson[201], L. M. Scott[215], P. D. Shah[201], F. Shibata[212], T. Shibata[208], H. Shimodaira[208], B. K. Shin[204], T. Shirahama[213], J. D. Smith[201], P. Sokolsky[201], R. W. Springer[201], B. T. Stokes[201], S. R. Stratton[201,215], T. A. Stroman[201], M. Takamura[205], M. Takeda[208], A. Taketa[226], M. Takita[208], Y. Tameda[211], H. Tanaka[210], K. Tanaka[227], M. Tanaka[220], S. B. Thomas[201], G. B. Thomson[201], P. Tinyakov[217,224], I. Tkachev[217], H. Tokuno[202], T. Tomida[228], S. Troitsky[217], Y. Tsunesada[202], K. Tsutsumi[202], Y. Uchihori[229], S. Udo[211], F. Urban[224], G. Vasiloff[201], Y. Wada[213], T. Wong[201], H. Yamaoka[220], K. Yamazaki[210], J. Yang[203], K. Yashiro[205], Y. Yoneda[210], S. Yoshida[219], H. Yoshii[230], R. Zollinger[201], Z. Zundel[201]

## The Pierre Auger Collaboration

A. Aab[40], P. Abreu[62], M. Aglietta[51], M. Ahlers[92], E.J. Ahn[80], I.F.M. Albuquerque[16], I. Allekotte[1], J. Allen[84], P. Allison[86], A. Almela[11,8], J. Alvarez Castillo[55], J. Alvarez-Muñiz[73], R. Alves Batista[39], M. Ambrosio[44], A. Aminaei[56], L. Anchordoqui[93], S. Andringa[62], T. Antičić[23], C. Aramo[44], F. Arqueros[70], H. Asorey[1], P. Assis[62], J. Aublin[29], M. Ave[73], M. Avenier[30], G. Avila[10], A.M. Badescu[66], K.B. Barber[12], R. Bardenet[28], J. Bäuml[34], C. Baus[36], J.J. Beatty[86], K.H. Becker[33], A. Bellétoile[32], J.A. Bellido[12], S. BenZvi[92], C. Berat[30], X. Bertou[1], P.L. Biermann[37], P. Billoir[29], F. Blanco[70], M. Blanco[29], C. Bleve[33], H. Blümer[36,34], M. Boháčová[25], D. Boncioli[50,45], C. Bonifazi[21], R. Bonino[51], N. Borodai[60], J. Brack[78], I. Brancus[63], P. Brogueira[62], W.C. Brown[79], P. Buchholz[40], A. Bueno[72], R.E. Burton[76], M. Buscemi[44], K.S. Caballero-Mora[73,87], B. Caccianiga[43], L. Caccianiga[29], M. Candusso[45], L. Caramete[37], R. Caruso[46], A. Castellina[51], G. Cataldi[48], L. Cazon[62], R. Cester[47], S.H. Cheng[87], A. Chiavassa[51], J.A. Chinellato[17], J. Chudoba[25], M. Cilmo[44], R.W. Clay[12], G. Cocciolo[48], R. Colalillo[44], L. Collica[43], M.R. Coluccia[48], R. Conceição[62], F. Contreras[9], H. Cook[74], M.J. Cooper[12], S. Coutu[87], C.E. Covault[76], A. Criss[87], J. Cronin[88], A. Curutiu[37], R. Dallier[32,31], B. Daniel[17], S. Dasso[5,3], K. Daumiller[34], B.R. Dawson[12], R.M. de Almeida[22], M. De Domenico[46], S.J. de Jong[56,58], G. De La Vega[7], W.J.M. de Mello Junior[17], J.R.T. de Mello Neto[21], I. De Mitri[48], V. de Souza[15], K.D. de Vries[57], L. del Peral[71], O. Deligny[27], H. Dembinski[34], N. Dhital[83], C. Di Giulio[45], J.C. Diaz[83], M.L. Díaz Castro[13], P.N. Diep[94], F. Diogo[62], C. Dobrigkeit [17], W. Docters[57], J.C. D'Olivo[55], P.N. Dong[94,27], A. Dorofeev[78], J.C. dos

Anjos[13], M.T. Dova[4], J. Ebr[25], R. Engel[34], M. Erdmann[38], C.O. Escobar[80,17], J. Espadanal[62], A. Etchegoyen[8,11], P. Facal San Luis[88], H. Falcke[56,59,58], K. Fang[88], G. Farrar[84], A.C. Fauth[17], N. Fazzini[80], A.P. Ferguson[76], B. Fick[83], J.M. Figueira[8,34], A. Filevich[8], A. Filipčič[67,68], N. Foerster[40], B.D. Fox[89], C.E. Fracchiolla[78], E.D. Fraenkel[57], O. Fratu[66], U. Fröhlich[40], B. Fuchs[36], R. Gaior[29], R.F. Gamarra[8], S. Gambetta[41], B. García[7], S.T. Garcia Roca[73], D. García-Gámez[28], D. Garcia-Pinto[70], G. Garilli[46], A. Gascon Bravo[72], H. Gemmeke[35], P.L. Ghia[29], M. Giller[61], J. Gitto[7], C. Glaser[38], H. Glass[80], F. Gomez Albarracin[4], M. Gómez Berisso[1], P.F. Gómez Vitale[10], P. Gonçalves[62], J.G. Gonzalez[36], B. Gookin[78], A. Gorgi[51], P. Gorham[89], P. Gouffon[16], S. Grebe[56,58], N. Griffith[86], A.F. Grillo[50], T.D. Grubb[12], Y. Guardincerri[3], F. Guarino[44], G.P. Guedes[18], P. Hansen[4], D. Harari[1], T.A. Harrison[12], J.L. Harton[78], A. Haungs[34], T. Hebbeker[38], D. Heck[34], A.E. Herve[12], G.C. Hill[12], C. Hojvat[80], N. Hollon[88], P. Homola[40,60], J.R. Hörandel[56,58], P. Horvath[26], M. Hrabovský[26,25], D. Huber[36], T. Huege[34], A. Insolia[46], P.G. Isar[64], S. Jansen[56,58], C. Jarne[4], M. Josebachuili[8,34], K. Kadija[23], O. Kambeitz[36], K.H. Kampert[33], P. Karhan[24], P. Kasper[80], I. Katkov[36], B. Kégl[28], B. Keilhauer[34], A. Keivani[82], E. Kemp[17], R.M. Kieckhafer[83], H.O. Klages[34], M. Kleifges[35], J. Kleinfeller[9,34], J. Knapp[74 d], R. Krause[38], N. Krohm[33], O. Krömer[35], D. Kruppke-Hansen[33], D. Kuempel[38], N. Kunka[35], G. La Rosa[49], D. LaHurd[76], L. Latronico[51], R. Lauer[91], M. Lauscher[38], P. Lautridou[32], S. Le Coz[30], M.S.A.B. Leão[14], D. Lebrun[30], P. Lebrun[80], M.A. Leigui de Oliveira[20], A. Letessier-Selvon[29], I. Lhenry-Yvon[27], K. Link[36], R. López[52], A. Lopez Agüera[73], K. Louedec[30], J. Lozano Bahilo[72], L. Lu[33,74], A. Lucero[8], M. Ludwig[36], H. Lyberis[21], M.C. Maccarone[49], C. Macolino[29], M. Malacari[12], S. Maldera[51], J. Maller[32], D. Mandat[25], P. Mantsch[80], A.G. Mariazzi[4], V. Marin[32], I.C. Mariş[29], H.R. Marquez Falcon[54], G. Marsella[48], D. Martello[48], L. Martin[32,31], H. Martinez[53], O. Martínez Bravo[52], D. Martraire[27], J.J. Masías Meza[3], H.J. Mathes[34], J. Matthews[82], J.A.J. Matthews[91], G. Matthiae[45], D. Maurel[34], D. Maurizio[13], E. Mayotte[77], P.O. Mazur[80], C. Medina[77], G. Medina-Tanco[55], M. Melissas[36], D. Melo[8], E. Menichetti[47], A. Menshikov[35], S. Messina[57], R. Meyhandan[89], S. Mićanović[23], M.I. Micheletti[6], L. Middendorf[38], I.A. Minaya[70], L. Miramonti[43], B. Mitrica[63], L. Molina-Bueno[72], S. Mollerach[1], M. Monasor[88], D. Monnier Ragaigne[28], F. Montanet[30], B. Morales[55], C. Morello[51], J.C. Moreno[4], M. Mostafá[78], C.A. Moura[20], M.A. Muller[17], G. Müller[38], M. Münchmeyer[29], R. Mussa[47], G. Navarra[51 ‡], J.L. Navarro[72], S. Navas[72], P. Necesal[25], L. Nellen[55], A. Nelles[56,58], J. Neuser[33], P.T. Nhung[94], M. Niechciol[40], L. Niemietz[33], T. Niggemann[38], D. Nitz[83], D. Nosek[24], L. Nožka[25], J. Oehlschläger[34], A. Olinto[88], M. Oliveira[62], M. Ortiz[70], N. Pacheco[71], D. Pakk Selmi-Dei[17], M. Palatka[25], J. Pallotta[2], N. Palmieri[36], G. Parente[73], A. Parra[73], S. Pastor[69], T. Paul[93,85], M. Pech[25], J. Pękala[60], R. Pelayo[52], I.M. Pepe[19], L. Perrone[48], R. Pesce[41], E. Petermann[90], S. Petrera[42], A. Petrolini[41], Y. Petrov[78], R. Piegaia[3], T. Pierog[34], P. Pieroni[3], M. Pimenta[62], V. Pirronello[46], M. Platino[8], M. Plum[38], M. Pontz[40], A. Porcelli[34], T. Preda[64], P. Privitera[88], M. Prouza[25], E.J. Quel[2], S. Querchfeld[33], S. Quinn[76], J. Rautenberg[33], O. Ravel[32], D. Ravignani[8], B. Revenu[32], J. Ridky[25], S. Riggi[49,73], M. Risse[40], P. Ristori[2], H. Rivera[43], V. Rizi[42], J. Roberts[84], W. Rodrigues de Carvalho[73], I. Rodriguez Cabo[73], G. Rodriguez Fernandez[45,73], J. Rodriguez Martino[9], J. Rodriguez Rojo[9], M.D. Rodríguez-Frías[71], G. Ros[71], J. Rosado[70], T. Rossler[26], M. Roth[34], B. Rouillé-d'Orfeuil[88], E. Roulet[1], A.C. Rovero[5], C. Rühle[35], S.J. Saffi[12], A. Saftoiu[63], F. Salamida[27], H. Salazar[52], F. Salesa Greus[78], G. Salina[45], F. Sánchez[8], P. Sanchez-Lucas[72], C.E. Santo[62], E. Santos[62], E.M. Santos[21], F. Sarazin[77], B. Sarkar[33], R. Sato[9], N. Scharf[38], V. Scherini[43], H. Schieler[34], P. Schiffer[39], A. Schmidt[35], O. Scholten[57], H. Schoorlemmer[89,56,58], P. Schovánek[25], F.G. Schröder[34,8], A. Schulz[34], J. Schulz[56], S.J. Sciutto[4], M. Scuderi[46], A. Segreto[49], M. Settimo[40,48], A. Shadkam[82], R.C. Shellard[13], I. Sidelnik[1], G. Sigl[39], O. Sima[65], A. Śmiałkowski[61], R. Šmída[34], G.R. Snow[90], P. Sommers[87], J. Sorokin[12], H. Spinka[75,80], R. Squartini[9], Y.N. Srivastava[85], S. Stanič[68], J. Stapleton[86], J. Stasielak[60], M. Stephan[38], M. Straub[38], A. Stutz[30], F. Suarez[8], T. Suomijärvi[27], A.D. Supanitsky[5], T. Šuša[23], M.S. Sutherland[82], J. Swain[85], Z. Szadkowski[61], M. Szuba[34], A. Tapia[8], M. Tartare[30], O. Taşcău[33], R. Tcaciuc[40], N.T. Thao[94], J. Tiffenberg[3], C. Timmermans[58,56], W. Tkaczyk[61 ‡], C.J. Todero Peixoto[15], G. Toma[63], L. Tomankova[34], B. Tomé[62], A. Tonachini[47], G. Torralba Elipe[73], D. Torres Machado[32], P. Travnicek[25], D.B. Tridapalli[16], E. Trovato[46], M. Tueros[73], R. Ulrich[34], M. Unger[34], J.F. Valdés Galicia[55], I. Valiño[73], L. Valore[44], G. van Aar[56], A.M. van den Berg[57], S. van Velzen[56],

A. van Vliet[39], E. Varela[52], B. Vargas Cárdenas[55], G. Varner[89], J.R. Vázquez[70], R.A. Vázquez[73], D. Veberič[68, 67], V. Verzi[45], J. Vicha[25], M. Videla[7], L. Villaseñor[54], H. Wahlberg[4], P. Wahrlich[12], O. Wainberg[8, 11], D. Walz[38], A.A. Watson[74], M. Weber[35], K. Weidenhaupt[38], A. Weindl[34], F. Werner[34], S. Westerhoff[92], B.J. Whelan[87], A. Widom[85], G. Wieczorek[61], L. Wiencke[77], B. Wilczyńska[60 ‡], H. Wilczyński[60], M. Will[34], C. Williams[88], T. Winchen[38], B. Wundheiler[8], S. Wykes[56], T. Yamamoto[88 a], T. Yapici[83], P. Younk[81], G. Yuan[82], A. Yushkov[73], B. Zamorano[72], E. Zas[73], D. Zavrtanik[68, 67], M. Zavrtanik[67, 68], I. Zaw[84 c], A. Zepeda[53 b], J. Zhou[88], Y. Zhu[35], M. Zimbres Silva[17], M. Ziolkowski[40]

[1] Centro Atómico Bariloche and Instituto Balseiro (CNEA-UNCuyo-CONICET), San Carlos de Bariloche, Argentina
[2] Centro de Investigaciones en Láseres y Aplicaciones, CITEDEF and CONICET, Argentina
[3] Departamento de Física, FCEyN, Universidad de Buenos Aires y CONICET, Argentina
[4] IFLP, Universidad Nacional de La Plata and CONICET, La Plata, Argentina
[5] Instituto de Astronomía y Física del Espacio (CONICET-UBA), Buenos Aires, Argentina
[6] Instituto de Física de Rosario (IFIR) - CONICET/U.N.R. and Facultad de Ciencias Bioquímicas y Farmacéuticas U.N.R., Rosario, Argentina
[7] Instituto de Tecnologías en Detección y Astropartículas (CNEA, CONICET, UNSAM), and National Technological University, Faculty Mendoza (CONICET/CNEA), Mendoza, Argentina
[8] Instituto de Tecnologías en Detección y Astropartículas (CNEA, CONICET, UNSAM), Buenos Aires, Argentina
[9] Observatorio Pierre Auger, Malargüe, Argentina
[10] Observatorio Pierre Auger and Comisión Nacional de Energía Atómica, Malargüe, Argentina
[11] Universidad Tecnológica Nacional - Facultad Regional Buenos Aires, Buenos Aires, Argentina
[12] University of Adelaide, Adelaide, S.A., Australia
[13] Centro Brasileiro de Pesquisas Fisicas, Rio de Janeiro, RJ, Brazil
[14] Faculdade Independente do Nordeste, Vitória da Conquista, Brazil
[15] Universidade de São Paulo, Instituto de Física, São Carlos, SP, Brazil
[16] Universidade de São Paulo, Instituto de Física, São Paulo, SP, Brazil
[17] Universidade Estadual de Campinas, IFGW, Campinas, SP, Brazil
[18] Universidade Estadual de Feira de Santana, Brazil
[19] Universidade Federal da Bahia, Salvador, BA, Brazil
[20] Universidade Federal do ABC, Santo André, SP, Brazil
[21] Universidade Federal do Rio de Janeiro, Instituto de Física, Rio de Janeiro, RJ, Brazil
[22] Universidade Federal Fluminense, EEIMVR, Volta Redonda, RJ, Brazil
[23] Rudjer Bošković Institute, 10000 Zagreb, Croatia
[24] Charles University, Faculty of Mathematics and Physics, Institute of Particle and Nuclear Physics, Prague, Czech Republic
[25] Institute of Physics of the Academy of Sciences of the Czech Republic, Prague, Czech Republic
[26] Palacky University, RCPTM, Olomouc, Czech Republic
[27] Institut de Physique Nucléaire d'Orsay (IPNO), Université Paris 11, CNRS-IN2P3, Orsay, France
[28] Laboratoire de l'Accélérateur Linéaire (LAL), Université Paris 11, CNRS-IN2P3, France
[29] Laboratoire de Physique Nucléaire et de Hautes Energies (LPNHE), Universités Paris 6 et Paris 7, CNRS-IN2P3, Paris, France
[30] Laboratoire de Physique Subatomique et de Cosmologie (LPSC), Université Joseph Fourier Grenoble, CNRS-IN2P3, Grenoble INP, France
[31] Station de Radioastronomie de Nançay, Observatoire de Paris, CNRS/INSU, France
[32] SUBATECH, École des Mines de Nantes, CNRS-IN2P3, Université de Nantes, France
[33] Bergische Universität Wuppertal, Wuppertal, Germany
[34] Karlsruhe Institute of Technology - Campus North - Institut für Kernphysik, Karlsruhe, Germany
[35] Karlsruhe Institute of Technology - Campus North - Institut für Prozessdatenverarbeitung und Elektronik, Karlsruhe, Germany
[36] Karlsruhe Institute of Technology - Campus South - Institut für Experimentelle Kernphysik (IEKP), Karlsruhe, Germany
[37] Max-Planck-Institut für Radioastronomie, Bonn, Germany
[38] RWTH Aachen University, III. Physikalisches Institut A, Aachen, Germany

[39] Universität Hamburg, Hamburg, Germany
[40] Universität Siegen, Siegen, Germany
[41] Dipartimento di Fisica dell'Università and INFN, Genova, Italy
[42] Università dell'Aquila and INFN, L'Aquila, Italy
[43] Università di Milano and Sezione INFN, Milan, Italy
[44] Università di Napoli "Federico II" and Sezione INFN, Napoli, Italy
[45] Università di Roma II "Tor Vergata" and Sezione INFN, Roma, Italy
[46] Università di Catania and Sezione INFN, Catania, Italy
[47] Università di Torino and Sezione INFN, Torino, Italy
[48] Dipartimento di Matematica e Fisica "E. De Giorgi" dell'Università del Salento and Sezione INFN, Lecce, Italy
[49] Istituto di Astrofisica Spaziale e Fisica Cosmica di Palermo (INAF), Palermo, Italy
[50] INFN, Laboratori Nazionali del Gran Sasso, Assergi (L'Aquila), Italy
[51] Osservatorio Astrofisico di Torino (INAF), Università di Torino and Sezione INFN, Torino, Italy
[52] Benemérita Universidad Autónoma de Puebla, Puebla, Mexico
[53] Centro de Investigación y de Estudios Avanzados del IPN (CINVESTAV), México, Mexico
[54] Universidad Michoacana de San Nicolas de Hidalgo, Morelia, Michoacan, Mexico
[55] Universidad Nacional Autonoma de Mexico, Mexico, D.F., Mexico
[56] IMAPP, Radboud University Nijmegen, Netherlands
[57] Kernfysisch Versneller Instituut, University of Groningen, Groningen, Netherlands
[58] Nikhef, Science Park, Amsterdam, Netherlands
[59] ASTRON, Dwingeloo, Netherlands
[60] Institute of Nuclear Physics PAN, Krakow, Poland
[61] University of Łódź, Łódź, Poland
[62] LIP and Instituto Superior Técnico, Technical University of Lisbon, Portugal
[63] 'Horia Hulubei' National Institute for Physics and Nuclear Engineering, Bucharest- Magurele, Romania
[64] Institute of Space Sciences, Bucharest, Romania
[65] University of Bucharest, Physics Department, Romania
[66] University Politehnica of Bucharest, Romania
[67] J. Stefan Institute, Ljubljana, Slovenia
[68] Laboratory for Astroparticle Physics, University of Nova Gorica, Slovenia
[69] Institut de Física Corpuscular, CSIC-Universitat de València, Valencia, Spain
[70] Universidad Complutense de Madrid, Madrid, Spain
[71] Universidad de Alcalá, Alcalá de Henares (Madrid), Spain
[72] Universidad de Granada and C.A.F.P.E., Granada, Spain
[73] Universidad de Santiago de Compostela, Spain
[74] School of Physics and Astronomy, University of Leeds, United Kingdom
[75] Argonne National Laboratory, Argonne, IL, USA
[76] Case Western Reserve University, Cleveland, OH, USA
[77] Colorado School of Mines, Golden, CO, USA
[78] Colorado State University, Fort Collins, CO, USA
[79] Colorado State University, Pueblo, CO, USA
[80] Fermilab, Batavia, IL, USA
[81] Los Alamos National Laboratory, Los Alamos, NM, USA
[82] Louisiana State University, Baton Rouge, LA, USA
[83] Michigan Technological University, Houghton, MI, USA
[84] New York University, New York, NY, USA
[85] Northeastern University, Boston, MA, USA
[86] Ohio State University, Columbus, OH, USA
[87] Pennsylvania State University, University Park, PA, USA
[88] University of Chicago, Enrico Fermi Institute, Chicago, IL, USA
[89] University of Hawaii, Honolulu, HI, USA
[90] University of Nebraska, Lincoln, NE, USA
[91] University of New Mexico, Albuquerque, NM, USA
[92] University of Wisconsin, Madison, WI, USA

[93] University of Wisconsin, Milwaukee, WI, USA
[94] Institute for Nuclear Science and Technology (INST), Hanoi, Vietnam

[201] High Energy Astrophysics Institute and Department of Physics and Astronomy, University of Utah, Salt Lake City, Utah, USA
[202] Graduate School of Science and Engineering, Tokyo Institute of Technology, Meguro, Tokyo, Japan
[203] Department of Physics and Institute for the Early Universe, Ewha Womans University, Seodaaemun-gu, Seoul, Korea
[204] Department of Physics and The Research Institute of Natural Science, Hanyang University, Seongdong-gu, Seoul, Korea
[205] Department of Physics, Tokyo University of Science, Noda, Chiba, Japan
[206] Department of Physics, Kinki University, Higashi Osaka, Osaka, Japan
[207] Department of Physics, Yonsei University, Seodaemun-gu, Seoul, Korea
[208] Institute for Cosmic Ray Research, University of Tokyo, Kashiwa, Chiba, Japan
[209] Kavli Institute for the Physics and Mathematics of the Universe (WPI), Todai Institutes for Advanced Study, the University of Tokyo, Kashiwa, Chiba, Japan
[210] Graduate School of Science, Osaka City University, Osaka, Osaka, Japan
[211] Faculty of Engineering, Kanagawa University, Yokohama, Kanagawa, Japan
[212] University of Yamanashi, Interdisciplinary Graduate School of Medicine and Engineering, Kofu, Yamanashi, Japan
[213] The Graduate School of Science and Engineering, Saitama University, Saitama, Saitama, Japan
[214] Astrophysical Big Bang Laboratory, RIKEN, Wako, Saitama, Japan
[215] Department of Physics and Astronomy, Rutgers University - The State University of New Jersey, Piscataway, New Jersey, USA
[216] Department of Physics, Tokyo City University, Setagaya-ku, Tokyo, Japan
[217] Institute for Nuclear Research of the Russian Academy of Sciences, Moscow, Russia
[218] Advanced Research Institute for Science and Engineering, Waseda University, Shinjuku-ku, Tokyo, Japan
[219] Department of Physics, Chiba University, Chiba, Chiba, Japan
[220] Institute of Particle and Nuclear Studies, KEK, Tsukuba, Ibaraki, Japan
[221] Faculty of Science, Kochi University, Kochi, Kochi, Japan
[222] Department of Physical Sciences, Ritsumeikan University, Kusatsu, Shiga, Japan
[223] Department of Physics, Sungkyunkwan University, Jang-an-gu, Suwon, Korea
[224] Service de Physique Théorique, Université Libre de Bruxelles, Brussels, Belgium
[225] Department of Astronomy and Space Science, Chungnam National University, Yuseong-gu, Daejeon, Korea
[226] Earthquake Research Institute, University of Tokyo, Bunkyo-ku, Tokyo, Japan
[227] Department of Physics, Hiroshima City University, Hiroshima, Hiroshima, Japan
[228] Advanced Science Institute, RIKEN, Wako, Saitama, Japan
[229] National Institute of Radiological Science, Chiba, Chiba, Japan
[230] Department of Physics, Ehime University, Matsuyama, Ehime, Japan

[‡] Deceased
[a] Now at Konan University
[b] Also at the Universidad Autonoma de Chiapas on leave of absence from Cinvestav
[c] Now at NYU Abu Dhabi
[d] now at DESY Zeuthen

# Progress Towards a Cross-Calibration of the Auger and Telescope Array Fluorescence Telescopes via an Air-borne Light Source

J.N. MATTHEWS [1] FOR THE PIERRE AUGER [2], AND TELESCOPE ARRAY [3], COLLABORATIONS

[1] *University of Utah, Department of Physics and Astronomy and High Energy Astrophysics Institute, Salt Lake City, UT USA*
[2] *Full Author List may be found at www.auger.org/archive/authors_2013_05.html*
[3] *Full Author List may be found at www.telescopearray.org/index.php/research/publications/conference-proceedings*

*jnm@physics.utah.edu*

**Abstract:** The optical calibration of the fluorescence telescopes is a significant contribution to the overall uncertainty of energy measurements made by the Pierre Auger Observatory and Telescope Array Project. Some sources of uncertainty, such as the fluorescence yield in air, affect both experiments similarly. However, the optical calibration of the fluorescence telescopes is a source of independent uncertainty. The Pierre Auger and Telescope Array collaborations have taken initial steps to establish a relative end-to-end optical calibration of the fluorescence telescopes. An Octocopter carrying a portable light source has been flown in front of fluorescence telescopes at both Pierre Auger and Telescope Array sites. Laboratory calibration measurements of the light source before and after the flights provide a common baseline for the relative end-to-end calibration. We expect this system will lead to a common photonic calibration for both experiments. After giving a brief description of the UV light source and the Octocopter used for the measurements, the parameters and the calibration procedures for the light source will be discussed. First results on telescope images of the light source will be presented.

## 1 Introduction

The Pierre Auger Observatory and Telescope Array Project experiments study cosmic rays at the highest energies. [1, 2] These experiments combine air fluorescence telescopes with a large array of surface detectors. The energy scale of both experiments is derived from the air fluorescence measurement which uses the Earth's atmosphere as a calorimeter. At present, Auger and Telescope Array appear to have an energy scale discrepancy of about 20%. [3] Some systematic uncertainties are known to affect the energy determination similarly for both experiments. However, one area where systematic uncertainties are expected to be largely independent is the photonic scale. It may be possible to reduce the apparent discrepancy between the experiments by comparing the response of the air fluorescence telescopes to a well understood and calibrated, portable light source. [4]

Such a source, a flying isotropic UV light source has been developed at Karlsruhe Institute of Technology (KIT). [5, 6] It has been in use at the Auger site in Argentina since 2010. [7] This portable light source is presently undergoing careful evaluation of isotropy, pulse rate and temperature dependence, and temporal stability.

The source is carried by a remote controlled flying platform called an Octocopter (Mikrokopter). [9, 10] The Octocopter is piloted via remote control and it includes GPS navigation and a magnetic compass, enabling it to fly to a pre-programmed position and maintain a stable position ($\sim$0.5m) and orientation ($\sim 5^{\circ}$) under favorable conditions. Additionally, pressure sensors are used to improve altitude accuracy and stabilization. Using this system we can accurately position a well understood light source within the field of view of both Auger and Telescope Array air fluorescence telescopes to compare their relative response.

## 2 The Octocopter

The Octocopter has its eight motors mounted in a circle with a diameter of 80cm. It has a maximum payload capacity of about 1kg. However, the weight of the payload significantly affects the available flight time. Since the calibration flights must be made in the dark of night when the fluorescence telescopes operate, LEDs were mounted on bottom of the arms supporting the motors. Green LEDs were installed on the arms in the forward direction and yellow LEDs were used on the remaining arms. See Figure 1. The navigation lights enable simple visual verification of position and orientation and can be switched on and off via remote control.

The Octocopter is typically flown to a distance of 1000m from a telescope where it maintains a stable position and orientation. When directed the light source emits a series of UV flashes at a rate of 1Hz. An on-board computer sends GPS information, temperature, optical pulse settings, and other data back to a base station via a wireless link where it is recorded for later analysis.

We have conducted three campaigns of flights. During the flights the on-board GPS recorded a typical positional stability of less than 1m under good flying conditions ($0.06^{\circ}$ at 1000m). The manufacturer data sheet indicates a systematic uncertainty in the absolute position of 2.5m. ($0.14^{\circ}$). We plan to independently evaluate the absolute positioning accuracy during future campaigns.

## 3 The Light Source

The flight time of the Octocopter depends strongly on the weight of the payload. Therefore, it is important that, in addition to being isotropic and stable, the source should be light-weight. Based on simulations and experiments, the light source was built using a twelve sided ABS plastic dodecahedron. The body was coated with Tyvek and a UV LED [11] was mounted at the center of each of the hexagonal faces. To further improve isotropy, the source is surrounded by a diffuser made of a thin spherical shell (r=50mm) of polystyrene etched with acetone. The total mass of the completed light source was less than 150g. The full system, including light source and batteries, allows for a maximum flight time of 15-20 min.

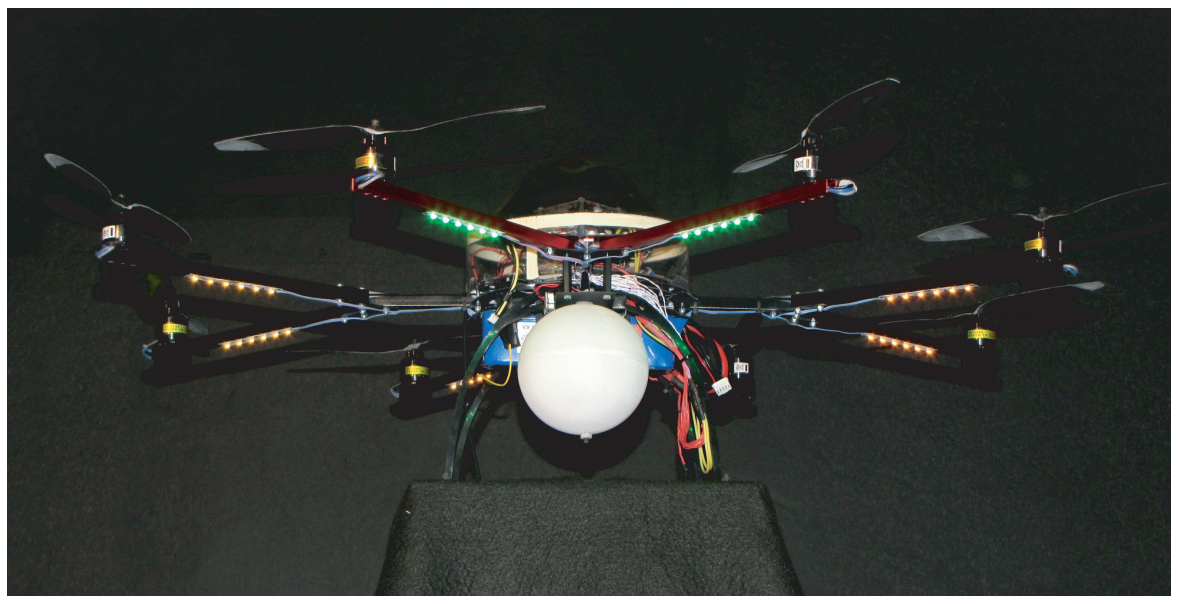

**Figure 1**: The Octocopter with the light source suspended below it. Green LEDs indicate the forward direction.

The UV LEDs provide 55mW of radiant flux at 350mA with a spectrum that peaks at 375nm with FWHM of about 10nm and a tail extending to 410nm. The LEDs emit light out to ($\sim \pm 90°$). However, they have a stronger emission peak in the forward direction (±10-15°) and a weaker broader peak (~80% of the strong peak out to $\sim \pm 60°$). Simulations of a similar source configuration which takes into account the mean light distribution pattern of the UV LEDs and their placement on the dodecahedron indicates deviations from isotropy should be of order 4% over the sphere. [6]

The current to each LED is individually controlled to compensate for variations in the LED intensities. There are six pre-programmed settings of pulse amplitude and variable width (2-64$\mu$sec) which can be used to illuminate the telescopes. The six standard settings span a factor of 10 in total intensity. The light pulse is triggered by the PPS signal of the GPS.

## 4 Source Calibration

A number of measurements have been performed to study effects that could influence the light intensity observed by the fluorescence telescopes. The main light source test bench is located at Karlsruhe. At KIT the spherical light source and a photodiode are mounted on an optical bench with a maximum separation of 2.5m. To block reflected light, one baffle (66mm diameter) and a black curtain are centered between them. The light from the sphere is measured with a NIST calibrated silicon photodiode, model UV100. NIST calibrates these detectors using a DC light source. [8] For the peak wavelength of the light source (375nm), NIST states a spectral power responsivity of 0.1293A/W with a relative combined uncertainty of 0.45% (for DC operation). The photodiode is equipped with a round baffle to reduce the active area to the homogeneous central region (0.5cm$^2$). The pulsed signal from the photodiode is then measured with a Keithley model 6514 electrometer that is readout via a USB connection.

The effect of temperature on the LED's output has been measured in the lab at KIT. The results, shown in Figure 2, show that the rate of change is dependent upon the driving amplitude for the LED. At present, this amplitude dependence is not well understood and studies are ongoing. At the highest setting, A5, the change is about -4% per 10°C. To allow one to make corrections for temperature dependence, the temperature inside the sphere and near the electronics are recorded.

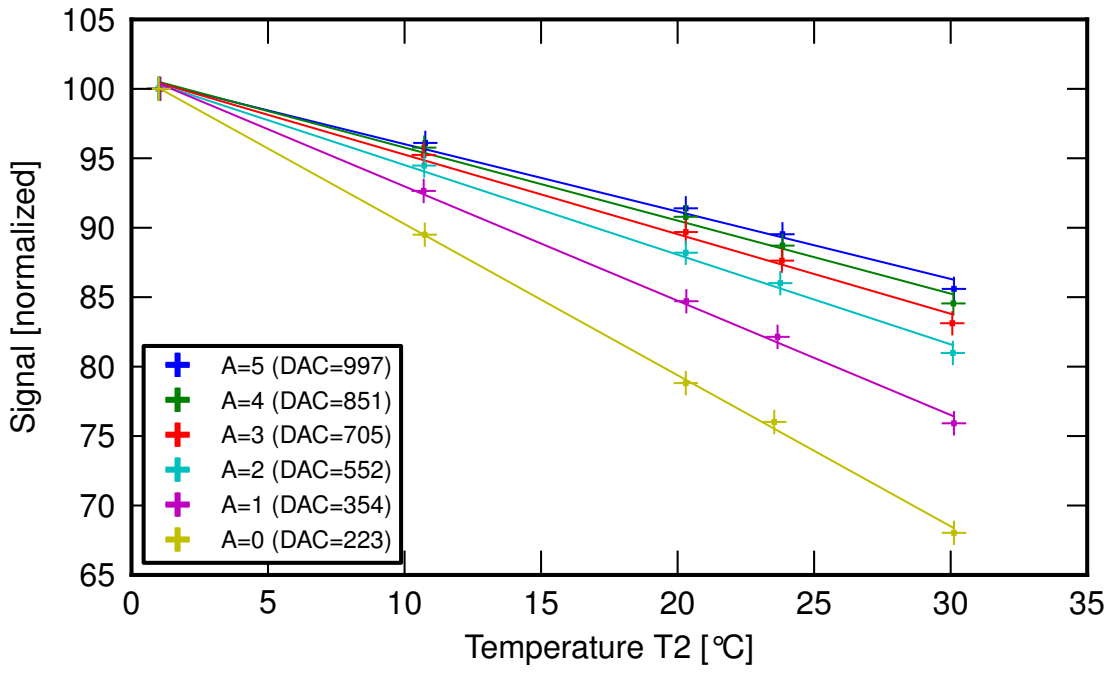

**Figure 2**: Relative signal strength from the Octocopter light source as a function of the temperature. Measurements were made at the DAC settings of the six standard amplitudes, A0 being the lowest and A5 the greatest. The amplitudes at each DAC setting are normalized to 100% at 0°C.

As the photodiode used for calibration of the source in the laboratory is considerably less sensitive than the air fluorescence telescopes, we must measure in the lab using a significantly higher pulse repetition rate of 1.1 kHz (900$\mu$sec period) *vs.* 1.00 Hz in the field. Initial studies of the dependence of the light output on the pulse rate suggest a correction of about -3.5% is required for field measurements. See Figure 3. The electrometer is configured to measure in 100msec intervals, each measurement triggers a pulse generator that in turn produces a burst of 100 triggers (90msec total) for the light source electronics. The 100 pulses are integrated to produce a measurement of total charge.

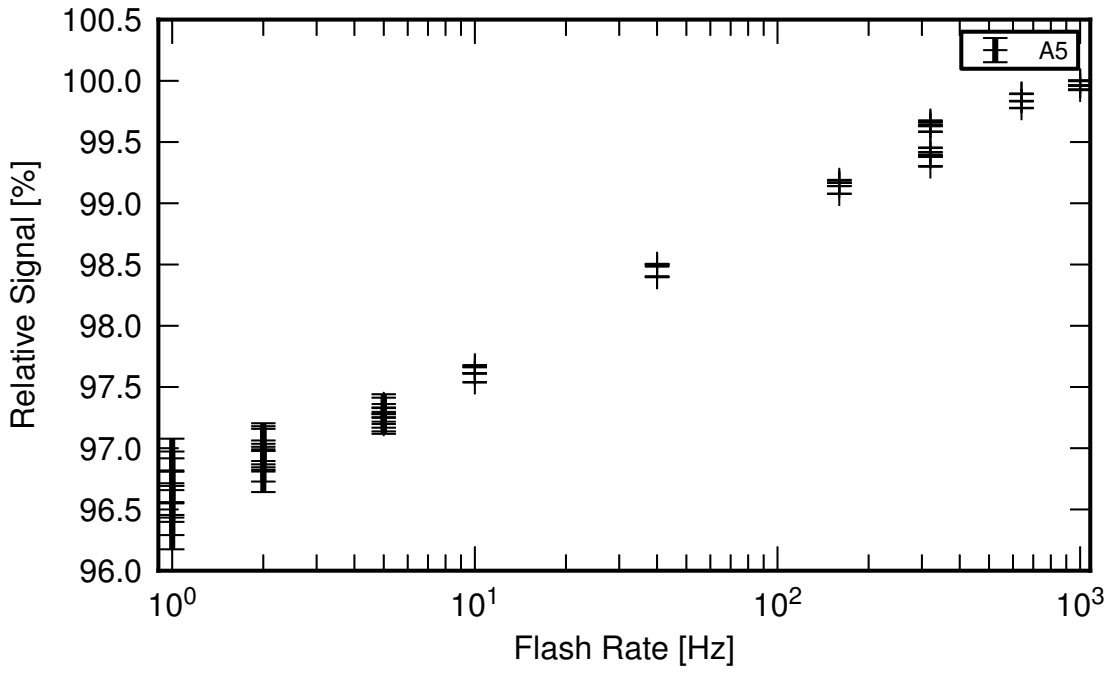

**Figure 3**: Relative signal strength from the Octocopter light source as a function of the flash rate. (Source amplitude setting 5)

Some tests of isotropy (in the forward direction) have been performed by rotating first in azimuth and then sepa-rately in elevation. For rotations less than $\pm 10°$ from forward in azimuth, the relative amplitude changed by about 0.2% (Figure 4). For elevation rotations, the relative amplitude changes by about 0.2% for a rotation of 10° and $<$3.5% for a rotation of 35° (Figure 5). This is consistent with the expectation from simulations noted earlier, however, more extensive evaluations are underway.

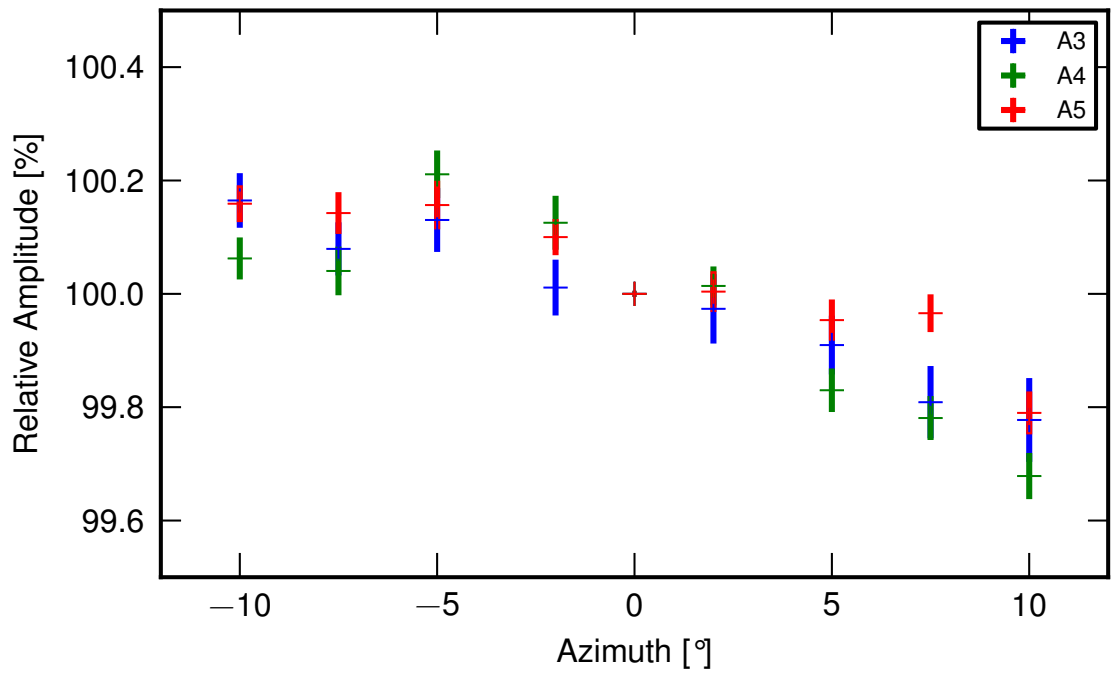

**Figure 4**: Relative signal strength from the Octocopter light source as a function of the azimuthal rotation angle from the forward direction. (Source amplitude settings A3(blue), A4(green), and A5(red))

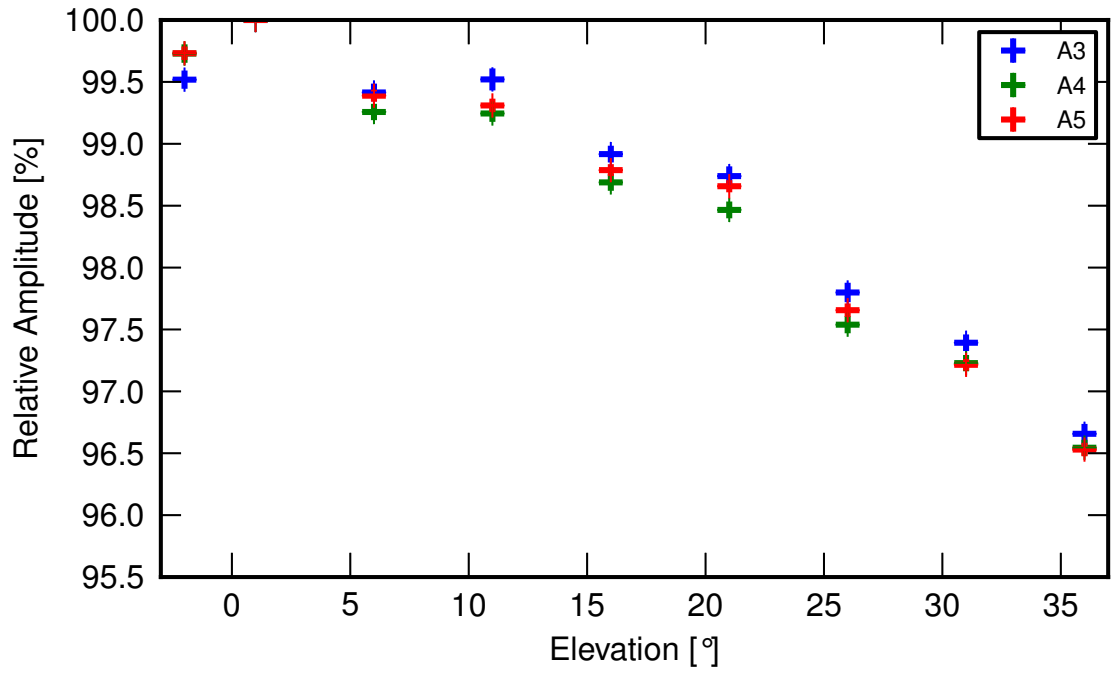

**Figure 5**: Relative signal strength from the Octocopter light source as a function of the elevation rotation angle from the forward direction. (Source amplitude settings A3(blue), A4(green), and A5(red))

To verify that any remaining reflections from the walls can be neglected, we checked the $1/r^2$ behavior of the setup for distances between 105cm and 245cm. The data was fitted with a $1/r^2$ function with constant background and a small, constant offset in $r$. The results showed that there is no significant contribution from the background light.

To monitor the stability of the light source, independent measurements of the light source were performed at the University of Utah before and after each of the two campaigns at the Telescope Array site in Delta, Utah. These lab measurements took place on 2012-10-11, 2012-10-18, 2013-03-12, and 2013-03-21. After correcting for temperature, the four Utah measurements, under stable laboratory conditions, agreed to $<$1% relative to the mean.

For the Utah measurements, the source was mounted inside a dark box at a fixed distance ($\sim$2.3m) from an NIST calibrated photodiode. It is important to note that this is a different photodiode from the one used for the measurements at Karlsruhe. Internal surfaces of the dark box were covered with low reflectance (black) cloth or flat black paint. In addition, a tube of black corrugated (egg carton) foam lined the optical path between the source and the photodiode to eliminate reflections. The NIST calibrated photodiode #G696 includes a precision calibrated aperture (50.12$\pm$0.05mm$^2$). The photodiode responsivity was measured at NIST in 5nm steps between 200nm and 1100nm. In the wavelength region near the peak of the source (375nm), the relative expanded uncertainty (k=2) of NIST photodiode #G696 responsivity is $<$1%. [8]

The signal from the photodiode was measured using a Keithley model 6485 picoampmeter. The picoampmeter recorded the mean current produced by series of flashes (period = 900$\mu$sec). As with the measurements at Karlsruhe, the system at Utah was configured to use a 100ms measurement interval. However, due to limitations of the hardware, instead of generating bursts of 100 triggers, an effectively continuous burst of 3000 triggers was generated. As a result, a single measurement cycle may contain either 111 or 112 flashes. This variation in the number of measured flashes contributes $\sim$0.1% to the measurement uncertainty. The temperatures inside the sphere and near the electronics were recorded to enable corrections for temperature dependence.

As a simple check of the charge measurement using the picoampmeter, a simple current source was constructed using a Lecroy 9210 Pulse Generator in combination with a 1.00G$\Omega$, 1% resistor. This enabled the generation of 8$\mu$sec duration current pulses with total charge in the lower range of that obtained from the light source and photodiode. Current pulses were generated and measured using the same software and identical settings to those used during the measurement of the light source. The measured charge per pulse for the simple current source agreed with the expected value to $\sim$1%.

Finally, after correcting for the temperature dependence of the source, measurements performed in the laboratories at Utah and Karlsruhe appear to agree to better than 2%. However, it is important to recognize that a number of potential sources of systematic uncertainty affecting measurements in the field remain under investigation.

## 5 Flights at the Auger Site

Octocopters have been flown at the Auger site in Argentina with various light sources for a series of tests since 2008. The Octocopter was flown at the Auger site with the current light source, as a part of this series of tests during a campaign in November 2012. The measurements took place shortly after the October series of flights at the Telescope Array site in Utah. This period of time was selected because weather conditions and temperatures could be expected to be similar for the northern (Telescope Array) and southern (Auger) hemisphere sites.

During the tests, the Octocopter flew inside the field of view of the 4th telescope located in the Los Leones building. This particular telescope has been extensively studied during previous Octocopter campaigns and its optical properties are believed to be well understood. The measurements took place during two nights 2012-11-5 and 2012-11-10 and included participants from the Telescope Array group.

Pixels in the center (column 10 and row 10), on the side (col. 17, row 10) and in the corner (col. 17, row 4) of the camera were illuminated from a distance of 1000 m with over 1100 flashes each. An event display from these flights is shown in Figure 6. Summing the signals over many flashes enables measurement of the point spread function down to a factor of $10^{-4}$ of the peak intensity. [7] The light source was also measured at KIT before and after this campaign. The observed change in intensity was <2%.

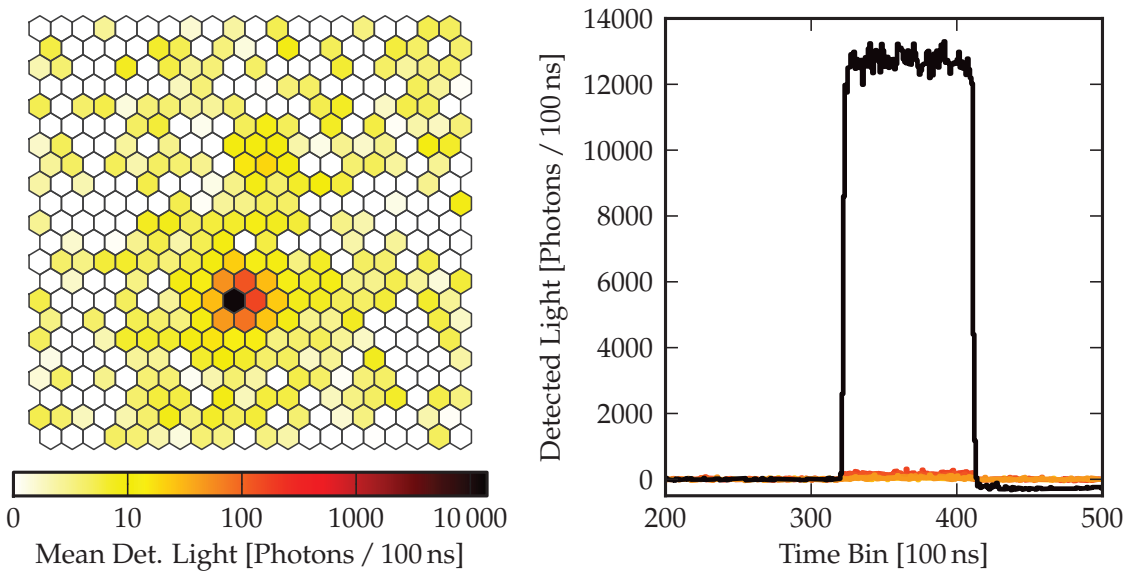

**Figure 6**: Event Display for an Octocopter light shot at the Auger Los Leones telescope site. The light level in the PMTs is shown at left, while the signal trace (photons/100ns) is shown at right.

## 6 Flights at the Telescope Array Site

In October 2012 and March 2013 a team from the Auger collaboration including KIT Octocopter experts visited the Telescope Array site in Utah to conduct campaigns of Octocopter light source flights. The team worked along with a researchers from the Telescope Array. The first test flights were conducted on 2012-10-14 at distances of 350m and 1000m from telescope 7 at the Black Rock Mesa site. The external temperature was about 0°C. Data was recorded using five of the six standard source settings. The Octocopter was positioned near the center of pixels 33, 73, 77, and B3 in telescope 7. Pixel 77 is located near the center of the camera and pixel 33 is half way to the corner. Pixel 33 would be expected to be more affected by spherical aberration. Pixels 73, 77, and B3 are equipped with a YAP to monitor the PMT response. [12, 13, 14] We note that pixel 77 in each camera is the telescope standard CRAYs calibrated PMT. [15] Additional flights were flown in the FOV of nine pixels the following two nights along with flights in pixel 77 at telescope 5. During these flights the highest light setting (A5) was used.

The Auger/KIT team returned to Utah for a second campaign in March, 2013. During this visit, there were only two nights of flights due to weather and conflicts with other tests. On 2013-03-16, the Octocopter flew in the Field Of View (FOV) of telescope 7 at the Black Rock Mesa site at a distance of 1000m. It hovered in the FOV of pixels 77, B7, 33, and B3. It also two flights with sweeps across the FOV of several PMTs. On 2013-03-19, it flew at a distance of 1000m in front of telescope 5 in the FOV of pixels 77 and 33.

## 7 Summary

Researchers at KIT have developed a sophisticated flying light source that can be used for *in-situ* optical calibrations of fluorescence telescopes. Measurement campaigns have been performed using this same equipment at the Telescope Array site (2012-10 and 2013-03) and at the Auger site (2012-11). Work is underway to understand and control systematics of the measurement to a level enabling productive comparisons between the Auger and Telescope Array photonic scales. Measurements of absolute light intensity, temperature dependence, rate dependence, and isotropy give us confidence these measurements will enable us to reduce the difference between Auger and Telescope Array photonic scales. The analysis of these datasets collected at the Auger and Telescope Array [16] is in progress and the results will be compared.

# Measuring Large-Scale Anisotropy in the Arrival Directions of Cosmic Rays Detected at the Telescope Array and the Pierre Auger Observatory above $10^{19}$ eV

Olivier Deligny[1] for the Telescope Array[2] and Pierre Auger[3] Collaborations.

[1] *IPN Orsay - CNRS/IN2P3 & Université Paris Sud*
[2] *Full author list may be found at http://www.telescopearray.org/index.php/research/publications/conference-proceedings*
[3] *Full author list may be found at http://www.auger.org/archive/authors_2013_05.html*

*ta-icrc@cosmic.utah.edu,auger_spokespersons@fnal.gov*

**Abstract:** Spherical harmonic moments are well-suited for characterizing anisotropy in the flux of cosmic rays. So far, above $10^{19}$ eV, no study has revealed a significant departure from isotropy. The dipole vector and the quadrupole tensor are of special interest, and access to the full set of multipoles could provide essential information for understanding the origin of ultra-high energy cosmic rays. Full-sky coverage allows the measurement of the spherical harmonic coefficients in an unambiguous way. This can be achieved by combining data from observatories located in both the northern and southern hemispheres. In this work, we present the prospects for a combined analysis using data recorded at the Telescope Array and the Pierre Auger Observatory. The main challenges are to account adequately for the relative exposures of both experiments and possibly different absolute energy normalizations. Using Monte-Carlo simulations, we show how these challenges will be addressed in an empirical way and illustrate the expected sensitivity of the methodology for the present observatory exposures.

The large-scale distribution of arrival directions of cosmic rays is an important observable in attempts to understand their origin. This is because this observable is closely connected to both their source distribution and their propagation. Due to the scattering in magnetic fields, the anisotropy imprinted in the arrival directions is mainly expected at large scales up to the highest energies. Large-scale patterns with anisotropy contrast at the level of $10^{-4}-10^{-3}$ have been reported by several experiments for energies below $\simeq 10^{15}$ eV where the high intensity of cosmic rays allows the collection of a large number of events. For energies above $\simeq 10^{15}$ eV, the decrease of the intensity with energy makes it more challenging to collect the statistics required to reveal amplitudes at the percent level which might be expected, in particular above $\simeq 10^{18}$ eV.

The anisotropy of any angular distribution on the sphere is encoded in the corresponding set of spherical harmonic moments $a_{\ell m}$. The dipole vector and the quadrupole tensor are of special interest, but an access to the full set of multipoles is relevant to characterise departures from isotropy at all scales. However, since cosmic ray observatories at ground level have only a partial-sky coverage, the recovering of these multipoles turns out to be nearly impossible without explicit assumptions on the shape of the angular distribution. In most cases, only the dipole (combination of $a_{1m}$ coefficients) and the quadrupole (combination of $a_{2m}$ coefficients) moments can be estimated with a sensible resolution under the assumption that the flux of cosmic rays is purely dipolar or purely dipolar and quadrupolar, respectively. Evading such hypotheses and thus measuring the multipoles to any order in an unambiguous way requires full-sky coverage. Full-sky coverage can only be achieved through the meta-analysis of data recorded by observatories located in both hemispheres of the Earth.

The Telescope Array, located in the Northern hemisphere (mean latitude $+39.3°$), and the Pierre Auger Observatory, located in the Southern hemisphere (mean latitude $-35.2°$), are the two largest experiments ever built dedicated to the study of ultra-high energy cosmic rays. Given the respective latitudes of both observatories, and given that the data sets from the Telescope Array and the Pierre Auger Observatory to be potentially combined consist of events with zenith angle up to $55°$ and $60°$ respectively, full-sky coverage can be indeed achieved. The present report is aimed at designing and studying in detail an effective way to combine data sets from both experiments while keeping under control the directional exposure. In the concern to facilitate this first joint analysis, the foreseen energy threshold, $10^{19}$ eV, is chosen to guarantee that both surface detector arrays operate with full detection efficiency for any of the local angles selected in each data set [1, 2]. This guarantees that each exposure function should follow purely geometric expectations to a high level. The main challenge in combining both data sets is to account adequately for the relative exposures of both experiments. In addition, since there are numerous sources of detector-dependent systematic uncertainties in the determination of the energy of a cosmic ray primary, there is presumably a difference in the energy scale between both experiments. Such a potential shift in energy leads to different counting rates above some fixed energy threshold, which induces fake anisotropies. Formally, these fake anisotropies are similar to the ones resulting from a shift in the relative exposures of the experiments, except in the case of energy-dependent anisotropies in the underlying flux of cosmic rays.

The observed angular distribution of cosmic rays, $dN/d\Omega$, can be naturally modeled as the sum of Dirac functions on the surface of the unit sphere whose arguments are the arrival directions $\{\mathbf{n}_1,...,\mathbf{n}_N\}$ of the events. Throughout the paper, arrival directions are expressed in the equatorial coordinate system (declination $\delta$ and right ascension $\alpha$) since this is the most natural one tied to the Earth to describe the directional exposure of any experiment. The random sample $\{\mathbf{n}_1,...,\mathbf{n}_N\}$ results from a Poisson process whose average is the flux of cosmic rays $\Phi(\mathbf{n})$ coupled to the directional exposure $\omega(\mathbf{n})$ of the considered experiment :

$$\left\langle \frac{dN(\mathbf{n})}{d\Omega} \right\rangle = \omega(\mathbf{n})\Phi(\mathbf{n}). \quad (1)$$

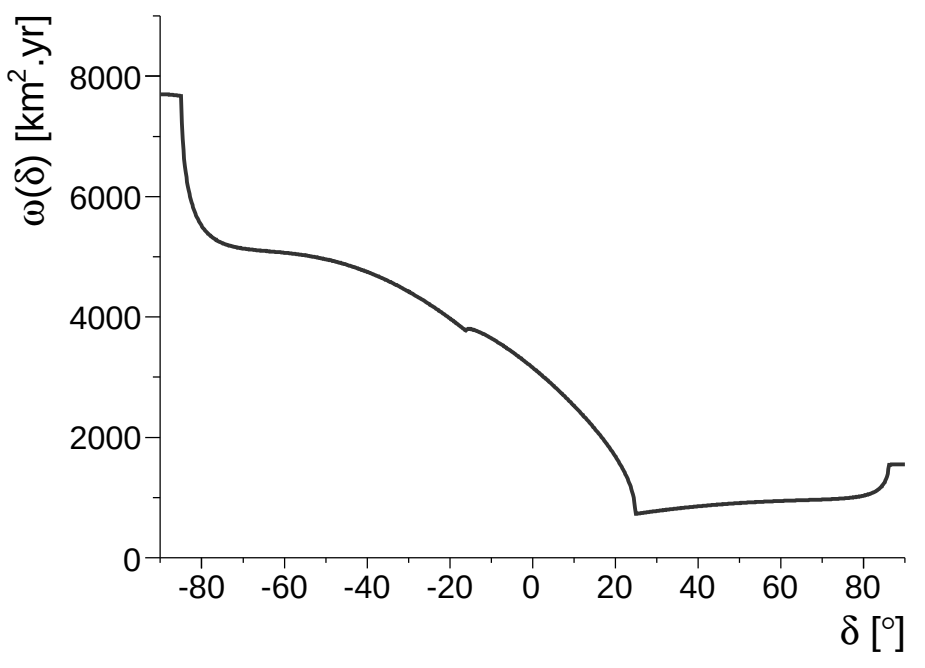

**Figure 1**: Total directional exposure above $10^{19}$ eV as obtained by summing the nominal individual ones of the Telescope Array and the Pierre Auger Observatory, as a function of the declination.

As any angular distribution on the unit sphere, the flux of cosmic rays $\Phi(\mathbf{n})$ can be decomposed in terms of a multipolar expansion onto the spherical harmonics $Y_{\ell m}(\mathbf{n})$ :

$$\Phi(\mathbf{n}) = \sum_{\ell \geq 0} \sum_{m=-\ell}^{\ell} a_{\ell m} Y_{\ell m}(\mathbf{n}). \quad (2)$$

Any anisotropy fingerprint is encoded in the $a_{\ell m}$ multipoles. Non-zero amplitudes in the $\ell$ modes arise from variations of the flux on an angular scale $\simeq 1/\ell$ radians.

The directional exposure of each observatory provides the effective time-integrated collecting area for a flux from each direction of the sky. In principle, the combined directional exposure of the two experiments should be simply the sum of the individual ones. However, individual exposures have here to be re-weighted by some empirical factor $b$ due to the unavoidable uncertainty in the relative exposures of the experiments. The parameter $b$ can be viewed as a fudge factor which absorbs any kind of systematic uncertainties in the relative exposures, whatever the sources of these uncertainties. This empirical factor is arbitrarily chosen to re-weight the directional exposure of the Pierre Auger Observatory relative to the one of the Telescope Array :

$$\omega(\mathbf{n};b) = \omega_{\mathrm{TA}}(\mathbf{n}) + b\,\omega_{\mathrm{Auger}}(\mathbf{n}). \quad (3)$$

Dead times of detectors modulate the directional exposure of each experiment in sidereal time and therefore in right ascension. However, once averaged over several years of data taking, the relative modulations of both $\omega_{\mathrm{TA}}$ and $\omega_{\mathrm{Auger}}$ in right ascension turn out to be not larger than few thousandths, yielding to non-uniformities in the observed angular distribution at the corresponding level. Given that the limited statistics currently available above $10^{19}$ eV cannot allow an estimation of each $a_{\ell m}$ coefficient with a precision better than a few percent, the non-uniformities of $\omega_{\mathrm{TA}}$ and $\omega_{\mathrm{Auger}}$ in right ascension can be neglected so that both functions are considered to depend only on the declination hereafter. On the other hand, since the high energy threshold guarantees that both experiments are fully efficient in their respective zenithal range $[0-\theta_{\mathrm{max}}]$, the dependence on declination is purely geometric [3] :

$$\omega_i(\mathbf{n}) = A_i \left( \cos\lambda_i \cos\delta \sin\alpha_m + \alpha_m \sin\lambda_i \sin\delta \right), \quad (4)$$

where $\lambda_i$ is the latitude of the considered experiment, the parameter $\alpha_m$ is given by

$$\alpha_m = \begin{cases} 0 & \text{if } \xi > 1, \\ \pi & \text{if } \xi < -1, \\ \arccos\xi & \text{otherwise}, \end{cases} \quad (5)$$

with $\xi \equiv (\cos\theta_{\mathrm{max}} - \sin\lambda_i \sin\delta)/\cos\lambda_i \cos\delta$, and the normalisation factors $A_i$ are tuned such that the integration of each $\omega_i$ function over $4\pi$ matches the (total) exposure of the corresponding experiment. For $b=1$, the resulting $\omega(\delta)$ function is shown in figure 1.

In practice, only an estimation $\overline{b}$ of the factor $b$ can be obtained, so that only an estimation of the directional exposure $\overline{\omega}(\mathbf{n}) \equiv \omega(\mathbf{n};\overline{b})$ can be achieved through equation 3. The procedure used for obtaining $\overline{b}$ from the joint data set will be described below. The resulting uncertainties propagate into uncertainties in the measured $\overline{a}_{\ell m}$ anisotropy parameters, in addition to the ones caused by the Poisson nature of the sampling process when the function $\omega$ is known exactly.

With full-sky but non-uniform coverage, the customary recipe for decoupling directional exposure effects from anisotropy ones consists in weighting the observed angular distribution by the inverse of the *relative* directional exposure function :

$$\frac{d\tilde{N}(\mathbf{n})}{d\Omega} = \frac{1}{\overline{\omega}_r(\mathbf{n})} \frac{dN(\mathbf{n})}{d\Omega}. \quad (6)$$

The relative directional exposure is the dimensionless function normalized to unity at its maximum. When the function $\omega$ (or $\omega_r$) is known from a single experiment, the averaged angular distribution $\langle d\tilde{N}/d\Omega \rangle$ is, from equation 1, identified with the flux of cosmic rays $\Phi(\mathrm{n})$ times the total exposure of the experiment. Due to the finite resolution to estimate $b$, the relationship between $\langle d\tilde{N}/d\Omega \rangle$ and $\Phi(\mathrm{n})$ is here not any longer so straightforward :

$$\left\langle \frac{d\tilde{N}(\mathbf{n})}{d\Omega} \right\rangle = \left\langle \frac{1}{\overline{\omega}_r(\mathbf{n})} \right\rangle \omega(\mathbf{n})\Phi(\mathbf{n}). \quad (7)$$

However, for an unbiased estimator of $b$ with a resolution better than $\simeq 10\%$ (the actual resolution on $b$ will be shown hereafter to be of the order of $\simeq 3.5\%$), the relative differences between $\langle 1/\overline{\omega}_r(\mathbf{n}) \rangle$ and $1/\omega_r(\mathbf{n})$ are actually smaller than $10^{-3}$ in such a way that $\langle d\tilde{N}/d\Omega \rangle$ can still be identified to $\Phi(\mathbf{n})$ times the total exposure to a high level. Consequently, the recovered $\overline{a}_{\ell m}$ coefficients defined as

$$\overline{a}_{\ell m} = \int_{4\pi} d\Omega \frac{d\tilde{N}(\mathbf{n})}{d\Omega} Y_{\ell m}(\mathbf{n}) = \sum_{i=1}^{N} \frac{Y_{\ell m}(\mathbf{n}_i)}{\overline{\omega}_r(\mathbf{n}_i)} \quad (8)$$

provide unbiased estimators of the underlying $a_{\ell m}$ multipoles since the relationship $\langle \overline{a}_{\ell m} \rangle = a_{\ell m}$ can be established by propagating equation 7 into $\langle \overline{a}_{\ell m} \rangle$.

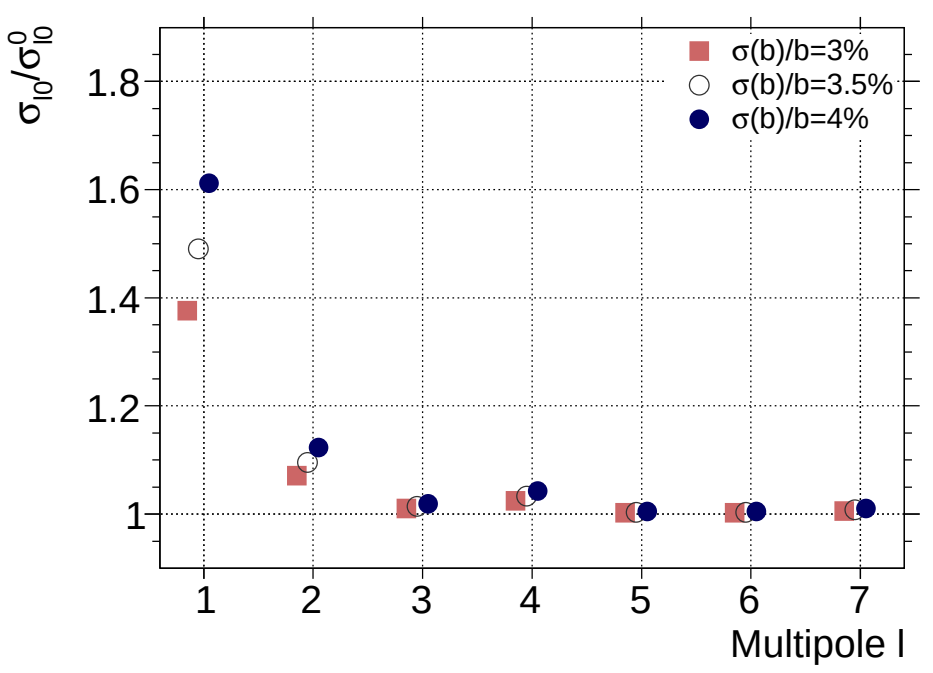

**Figure 2**: Illustration of the dependence of the resolution on the recovered $\overline{a}_{\ell 0}$ coefficients upon the uncertainty on $b$, for different values of the resolution on $b$. On the $y-$axis, the term $\sigma^0_{\ell 0}$ is obtained by dropping the second term inside the square root in the expression of $\sigma_{\ell 0}$ (see equation 9).

Using the estimators defined in equation 8, the expected resolution $\sigma_{\ell m}$ on each $a_{\ell m}$ multipole can be inferred by propagating the second moment of $d\tilde{N}/d\Omega$ into the covariance matrix of the estimated $\overline{a}_{\ell m}$ coefficients accordingly to the Poisson statistics. In the case of relatively small $\{a_{\ell m}\}_{\ell \geq 1}$ coefficients compared to $a_{00}$, this leads to :

$$\sigma_{\ell m} \simeq \frac{a_{00}}{\sqrt{4\pi}} \left[ \frac{\sqrt{4\pi}}{a_{00}} \int_{4\pi} d\Omega \left\langle \frac{1}{\overline{\omega}_r^2(\mathbf{n})} \right\rangle \omega(\mathbf{n}) Y^2_{\ell m}(\mathbf{n}) + \int_{4\pi} d\Omega d\Omega' \left[ \left\langle \frac{1}{\overline{\omega}_r(\mathbf{n})\overline{\omega}_r(\mathbf{n}')} \right\rangle \omega(\mathbf{n})\omega(\mathbf{n}') - 1 \right] Y_{\ell m}(\mathbf{n}) Y_{\ell m}(\mathbf{n}') \right]^{1/2} . \tag{9}$$

If $b$ were known with perfect accuracy, the second term in equation 9 would vanish, and the resolution of the $a_{\ell m}$'s would be similar to that for a single experiment. The second term adds the effect of the uncertainty in the relative exposures of the two experiments. For a directional exposure independent of the right ascension, it is non-zero for $m = 0$ only, as expected. Its influence is illustrated in figure 2, where the ratio between the total expression of $\sigma_{\ell 0}$ and the partial one ignoring this second term inside the square root is plotted as a function of the multipole $\ell$ for different resolution values on $b$. While this ratio amounts to $\simeq 1.5$ for $\ell = 1$ and $\sigma(b)/b = 3.5\%$, it falls to $\simeq 1.1$ for $\ell = 2$ and then tends to 1 for higher multipole values. Consequently, in accordance with naive expectations, the uncertainty on the $b$ factor mainly impacts the resolution on the dipole coefficient $a_{10}$ while it has a small influence on the quadrupole coefficient $a_{20}$ and a marginal one on higher order moments $\{a_{\ell 0}\}_{\ell \geq 3}$.

The hybrid nature of both observatories enables the assignation of the energy of each event to be derived in a calorimetric way through the calibration of the shower size measured with the SD arrays by the energy measured with the fluorescence telescopes on a subset of high quality hybrid events [4, 5]. Nevertheless, though the techniques are nearly the same, there are differences as to how the primary energies are derived at the Telescope Array and the Pierre Auger Observatory. Currently, systematic uncertainties in the energy scale of both experiments amount to about 20% and 14% respectively [6, 7]. Uncovering and understanding the sources of systematic uncertainties in the respective energy scales is out of the scope of this report and will be addressed elsewhere (see for instance [8]). Rather, the aim pursued here is only to guarantee that the relative exposures between both observatories is kept under control in an accurate way, whatever the unknown differences in energy scale. To this end, a purely empirical cross-calibration procedure exploiting the overlapping part of the sky exposed to both experiments is designed for estimating *in fine* reliable anisotropy parameters.

A band of declinations around the equatorial plane is exposed to the fields of view of both experiments, namely for declinations between $-15^\circ$ and $25^\circ$. This overlapping region can be used for *cross-calibrating empirically* the energy scales and for delivering an overall unbiased estimation of the $a_{\ell m}$ multipoles in the case of isotropy. Though the cross-calibration of the energy scale is not a mandatory input for the procedure, it constitutes however a reasonable starting point for studying anisotropies in the arrival directions of all events detected in excess of roughly the same energy threshold by both experiments. The procedure is based on an iterative algorithm which is now detailed. Considering as a first approximation the flux $\Phi(\mathbf{n})$ as isotropic, and given $N_{\mathrm{TA}}$ events observed in total above $10^{19}$ eV, the number of events $N_{\mathrm{Auger}}$ corresponding to that particular energy threshold can be inferred accordingly to a simple proportionality :

$$N_{\mathrm{Auger}} = \frac{\int_{4\pi} d\Omega\ \omega_{\mathrm{Auger}}(\mathbf{n})}{\int_{4\pi} d\Omega\ \omega_{\mathrm{TA}}(\mathbf{n})} N_{\mathrm{TA}}. \tag{10}$$

The energy threshold guaranteeing equal intensities for both experiments is then provided by selecting the $N_{\mathrm{Auger}}$ highest energy events to be combined with the $N_{\mathrm{TA}}$ events. The resulting joint data set consists then of all events with energies in excess of $10^{19}$ eV in terms of the energy scale used at the Telescope Array. Using the joint data set built in this way, a first estimation of $b$ can be made by counting the $\Delta N_{\mathrm{TA}}$ and $\Delta N_{\mathrm{Auger}}$ observed in the overlapping region $\Delta\Omega$ :

$$\overline{b}^{(0)} = \frac{\Delta N_{\mathrm{Auger}}}{\Delta N_{\mathrm{TA}}} \frac{\int_{\Delta\Omega} d\Omega\ \omega_{\mathrm{TA}}(\mathbf{n})}{\int_{\Delta\Omega} d\Omega\ \omega_{\mathrm{Auger}}(\mathbf{n})}. \tag{11}$$

Inserting $\overline{b}^{(0)}$ into $\overline{\omega}$, 'zero-order' $\overline{a}^{(0)}_{\ell m}$ coefficients can be obtained. This set of coefficients is only a rough estimation, due to the limiting assumption on the flux (isotropy).

On the other hand, the expected number of events in the common band for each observatory, $\Delta n^{\mathrm{exp}}_{\mathrm{TA}}$ and $\Delta n^{\mathrm{exp}}_{\mathrm{Auger}}$, can be expressed from the underlying flux $\Phi(\mathbf{n})$ and the true value of $b$ as :

$$\begin{aligned} \Delta n^{\mathrm{exp}}_{\mathrm{TA}} &= \int_{\Delta\Omega} d\Omega\ \Phi(\mathbf{n})\omega_{\mathrm{TA}}(\mathbf{n}) \\ \Delta n^{\mathrm{exp}}_{\mathrm{Auger}} &= b \int_{\Delta\Omega} d\Omega\ \Phi(\mathbf{n})\omega_{\mathrm{Auger}}(\mathbf{n}). \end{aligned} \tag{12}$$

From equations 12, and from the set of $\overline{a}^{(0)}_{\ell m}$ coefficients, an iterative procedure estimating at the same time $b$ and the set of $a_{\ell m}$ coefficients can be considered in the following way :

$$\overline{b}^{(k+1)} = \frac{\Delta N_{\mathrm{Auger}}}{\Delta N_{\mathrm{TA}}} \frac{\int_{\Delta\Omega} d\Omega\ \overline{\Phi}^{(k)}(\mathbf{n})\omega_{\mathrm{TA}}(\mathbf{n})}{\int_{\Delta\Omega} d\Omega\ \overline{\Phi}^{(k)}(\mathbf{n})\omega_{\mathrm{Auger}}(\mathbf{n})}, \tag{13}$$

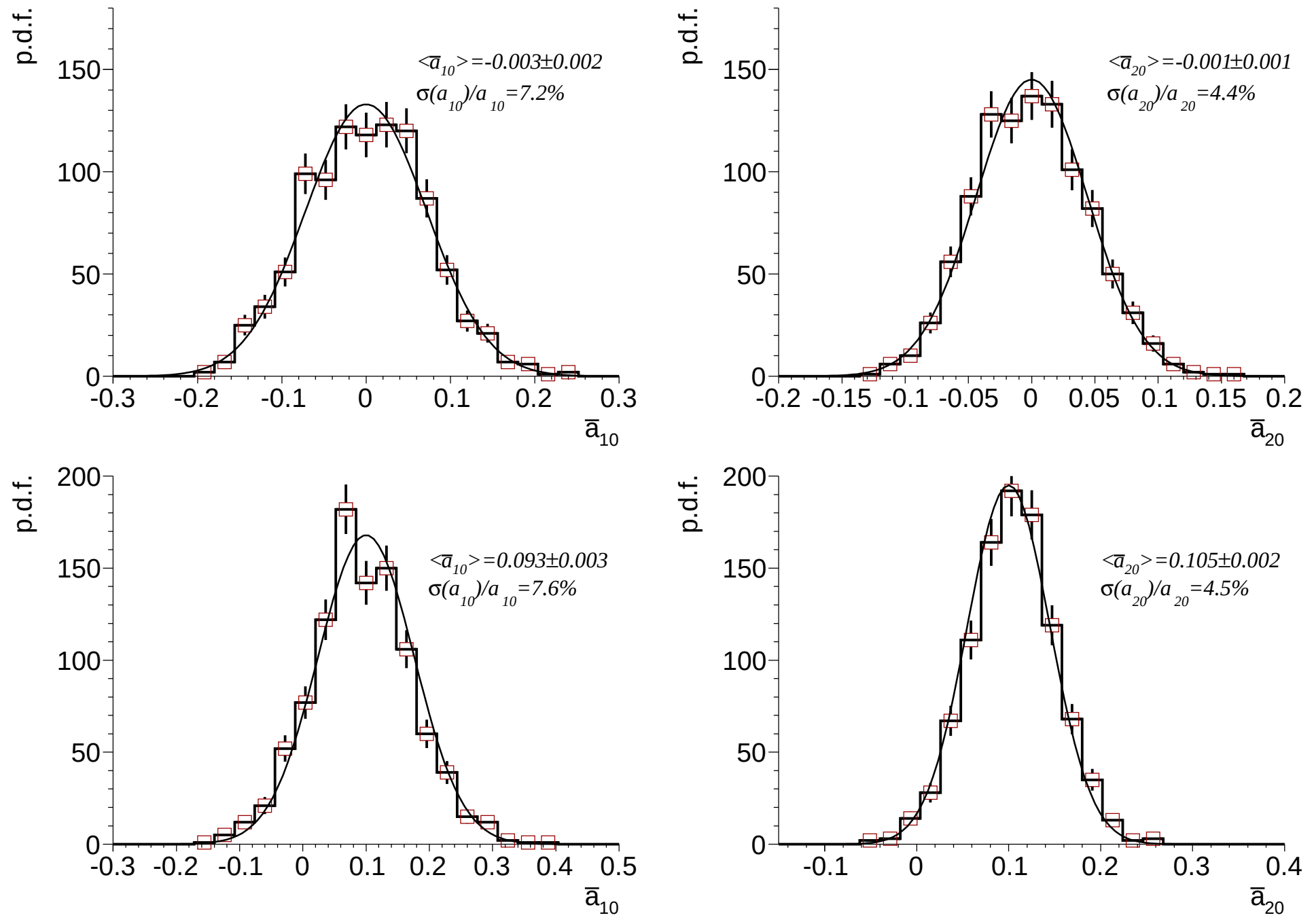

**Figure 3**: Reconstruction of $a_{10}$ (left) and $a_{20}$ (right) through the iterative procedure, in the case of an underlying isotropic flux (top) or of an anisotropic input flux $\Phi(\mathbf{n}) \propto 1+0.1Y_{10}(\mathbf{n})+0.1Y_{20}(\mathbf{n})$ (bottom). Expectations are shown as the Gaussian curves with resolution parameters as in equation 9.

where $\Delta N_{\mathrm{Auger}}$ and $\Delta N_{\mathrm{TA}}$ as derived in the first step are used to estimate $\Delta n_{\mathrm{TA}}^{\mathrm{exp}}$ and $\Delta n_{\mathrm{Auger}}^{\mathrm{exp}}$ respectively, and $\overline{\Phi}^{(k)}$ is the flux estimated with the set of $\overline{a}_{\ell m}^{(k)}$ coefficients.

Whether this iterative procedure delivers *in fine* unbiased estimations of the set of $a_{\ell m}$ coefficients with a resolution given by equation 9 can be tested by Monte-Carlo simulations. For 1,000 mock samples with a number of events similar to the one of the actual joint data set and with ingredients corresponding to the actual figures in terms of total and directional exposures, the distributions of reconstructed low orders $\overline{a}_{10}$ and $\overline{a}_{20}$ multipoles are shown in figure 3 (top panels) after $k=10$ iterations in the case of an underlying isotropic flux of cosmic rays. A systematic shift of 20% in energy scale is simulated in each sample; while the directional exposures used in equation 10 correspond to the ones used for generating the events. The averages of the reconstructed histograms are found in agreement with expectations; while, taking as input the observed RMS of the distribution of $\overline{b}$ ($\simeq 3.5\%$) and assuming Gaussian functions, the RMS of the $\overline{a}_{\ell m}$ distributions are found in agreement with equation 9. In practice, these results are found to be stable as soon as $k=4$.

With the exactly same ingredients, the simulations can be used to test the procedure with an underlying anisotropic flux of cosmic rays, chosen here such that $\Phi(\mathbf{n}) \propto 1+0.1Y_{10}(\mathbf{n})+0.1Y_{20}(\mathbf{n})$. Results of the Monte-Carlo simulations are shown in figure 3 (bottom panels) for the specific $a_{10}$ and $a_{20}$ coefficients. In the case of an underlying anisotropic flux of cosmic rays, it is important to note that deciding upon a fixed number of events $N_{\mathrm{Auger}}$ through an equation valid in the case of isotropy only (equation 10) is expected to imprint some fake pattern that cannot be fully absorbed. The resulting biases on the $\overline{a}_{10}$ and $\overline{a}_{20}$ are however small, as evidenced in figure 3 (bottom panels).

The cross-calibration procedure designed in this study makes it possible to use the powerful multipolar analysis method for characterising the sky map of ultra-high energy cosmic rays. It pertains to any full-sky coverage achieved by combining data sets from different observatories, and opens a rich field of anisotropy studies. This technique will be applied to data sets from the Telescope Array and the Pierre Auger Observatory and will be reported in a near future.

# Progress towards understanding the analyses of mass composition made by the Auger and Telescope Array Collaborations

WILLIAM F. HANLON[1] FOR THE TELESCOPE ARRAY[2] AND PIERRE AUGER COLLABORATIONS[3].

[1] *University of Utah, Department of Physics and Astronomy and High Energy Astrophysics Institute, Salt Lake City, UT, USA*
[2] *http://www.telescopearray.org/index.php/research/publications/conference-proceedings*
[3] *Full author list: http://www.auger.org/archive/authors_2013 05.html*

*whanlon@cosmic.utah.edu*

**Abstract:** The composition of cosmic rays at the highest energies is one of the most important problems in UHE cosmic ray physics. Recent results using fluorescence and hybrid fluorescence/surface array detectors (HiRes/Telescope Array/Auger) appear to lead to inconsistent conclusions. Comparison is not straightforward because of different acceptance and resolution of the various experiments. Here we take a 4-component mixture of protons, helium, nitrogen, and iron that varies with energy in such a way that it reproduces the Auger $X_{\max}$ data (Auger $X_{\max}$ data as obtained with hybrid measurements). We use this mix to simulate air showers in the TA aperture. These events are then passed through the TA detector simulation and reconstructed using TA hybrid methods and cuts. In this paper we describe the method and present the results of the simulations. The results show that the $\langle X_{\max} \rangle$ for the Auger mix composition would be observed by TA hybrid (after full event reconstruction) with a bias of $5.2 \pm 0.4$ g/cm$^2$, and the pure proton coposition will be observe with a bias of $11.5 \pm 0.9$ g/cm$^2$. The difference in the expected $\langle X_{\max} \rangle$ (reconstructed by TA-hybrid) between the pure proton and the Auger mix compositions is 20 g/cm$^2$ at $10^{19}$ eV, and the present study shows that, given the number of events generated, the Telescope Array would be able to distinguish between these two compositions with a confidence level better than 4 sigmas.

## 1 Introduction

One of the most important goals in particle astrophysics is understanding the chemical composition of ultra high energy cosmic rays (UHECRs). Knowledge of the relative proportions of UHECR species arriving at the earth will constrain models of cosmic ray origin and propagation, which are currently controversial (see [2] for example). Measurement of the UHECR composition in a event-by-event basis at the highest energies is difficult (due to the extremely low flux). Therefore the composition must be inferred indirectly by measuring the depth of shower maximum ($X_{\max}$) via fluorescence detection.

For a given shower, the depth of shower maximum depends upon the depth of first interaction ($X_0$), which decreases with $\log(E_0)$, and the depth over which the shower cascade takes to develop until the mean energy per secondary particle falls below the critical energy at which collision losses exceed radiative losses. Though all of the details needed to model ultra high energy air showers are not completely understood (cross sections, multiplicities, etc.), [1] describes a simple branching model of air shower development, introduced by Heitler, which reveals two important characteristics of the air showers: the $\langle X_{\max} \rangle$ is proportional $\ln(E_0)$ (where $E_0$ is the primary particle energy) and the elongation rate is constant for a given primary particle composition. The elongation rate is defined as $\mathrm{d}\langle X_{\max} \rangle / \mathrm{d}\log E$, which is the change of the mean $X_{\max}$ per decade of primary particle energy.

The Heitler model can be extended to showers initiated by nuclei of any given atomic number $A$ by invoking the superposition principle. In this case we can treat the shower as $A$ primary showers each with initial energy $E_0/A$. Showers initiated by heavier nuclei develop faster (i.e., $X_{\max}$ will be smaller). The $X_{\max}$ shower-to-shower fluctuations are not described by the simplistic superposition model, because it does not take into account effects from nuclear fragmentation, impact parameter fluctuations, etc. However, the $X_{\max}$ fluctuations are expected to reduced for larger $A$ due to averaging effects. We therefore expect the width of the distribution of $X_{\max}$ to be sensitive to the primary particle as well.

The distribution of $X_{\max}$ observed in a given energy bin is dependent upon the statistical fluctuations of shower development (depth of first interaction and cascade development) in the atmosphere as well as upon the resolution of the detector. Using the Heitler model we expect a spectrum composed of light particles (e.g. protons) to have larger mean $X_{\max}$ and a distribution width larger than that of a heavier species (e.g. iron). Additionally if the composition is unchanging over different energy ranges the elongation rate will remain constant.

Telescope Array (succeeding the HiRes experiment) with 750 km$^2$ of collecting area, described in [3] and [4], and the Pierre Auger Observatory with 3000 km$^2$ of collecting area, described in [5], are the 2 largest cosmic ray observatories. Both deploy large surface arrays to detect charged particles (and protons in the case of Auger) which reach the Earth's surface, as well as multiple fluorescence telescopes placed around the array, to observe UV light caused by the electromagnetic cascade of the air shower. While $X_{\max}$ is determined by using the shower profile as observed by the fluorescence detectors, folding in the geometry and timing information of the surface detectors for those showers that trigger them can improve the profile fit and further constrain $X_{max}$. In [6], HiRes reported measuring a lighter composition composed primarily of protons above $10^{18.2}$ eV. The Pierre Auger collaboration reported in [7] measurements of the first two moments of the $X_{max}$ distribution, the mean and the RMS, above $10^{18}$ eV with narrowing $X_{max}$ widths above $3 \times 10^{18.5}$ eV, indicating a composition possibly changing from lighter to heavier species.

At the UHECR 2012 conference in Geneva, Switzerland in March 2012 the Auger and Telescope Array (TA) collaborations formed a Mass Composition Working Group (MCWG) to discuss how the two groups could work together to resolve outstanding differences in the interpretation of conflicting $X_{max}$ data [8]. It was decided that Auger would provide simulated data which resembles the Auger $X_{max}$ distributions. This simulated data would be reconstructed through the TA analysis software to examine the effects of reconstruction of Auger $X_{max}$ input with TA detector effects folded in and then compare those results with observed $X_{max}$ as seen in TA data. In particular, the question of whether TA detector resolution and number of events would prevent TA from seeing a changing composition or a composition that is heavier than protons at the highest energies, could be addressed.

## 2 Data Analysis

Auger created an *ad hoc* model of UHECR composition by examining their data published in [9], and fitting it with a 4-component mixture. The model is called *ad hoc* because it is not claimed to be a physical model of what the actual cosmic ray beam contains. Using a reasonable choice of 4 input species, proton, helium, nitrogen, and iron the best proportions were found by fitting the expected $X_{max}$ distributions to the Auger $X_{max}$ data distribution. The four species fractions were found for each energy bin. Figure 1 shows the $\chi^2$ fits Auger performed on their $X_{max}$ data using a 4-component model. The 4-component fractions found using that fit was used to generate the Monte Carlo studied in this paper. The left figure compares the fit (red points) to the data (black points) $\langle X_{max} \rangle$ and the right compares the fit and $X_{max}$ widths. There is a $\sim 8$ g/cm$^2$ bias between the means from the fit and the data which is caused by low statistics in the tails of the $X_{max}$ distributions. A maximum likelihood fit was later performed reducing the bias in the means and is shown in figure 2. However we used the fractions of the 4-component mixture found from the $\chi^2$ fits in the present analysis.

TA generated a Monte Carlo set of $\sim 4$ million events using the 4 input species in the same relative proportions as described by the Auger mixture weighted to the HiRes1 and HiRes2 combined mono spectra shown in [10] in 0.1 decadal bins. Events that triggered the surface detector array and at least 1 of the 2 fluorescence detectors ("hybrid" events) were accepted for analysis. A TA surface array trigger consists of 3 SDs counters, above 3 MIP each, within an 8 microsecond window (as described in [3]). A TA FD trigger consists of at least 5 adjoining PMTs above night sky background within a coincidence window of 25.6 microseconds (as described in [11]).

Auger claims to reconstruct shower $X_{max}$ with very little bias due to fiducial volume cuts based on each shower's geometry. We expect then that Auger data should closely resemble the $\langle X_{max} \rangle$ from the input (thrown) Monte Carlo. Figure 3 compares the $\langle X_{max} \rangle$ of the Auger data described in [9] with the thrown $\langle X_{max} \rangle$ of the composition mixture after the Telescope Array SD trigger bias and the agreement is very good, indicating little bias in the $\langle X_{max} \rangle$ between the thrown Monte Carlo tested by Telescope Array and the real Auger data.

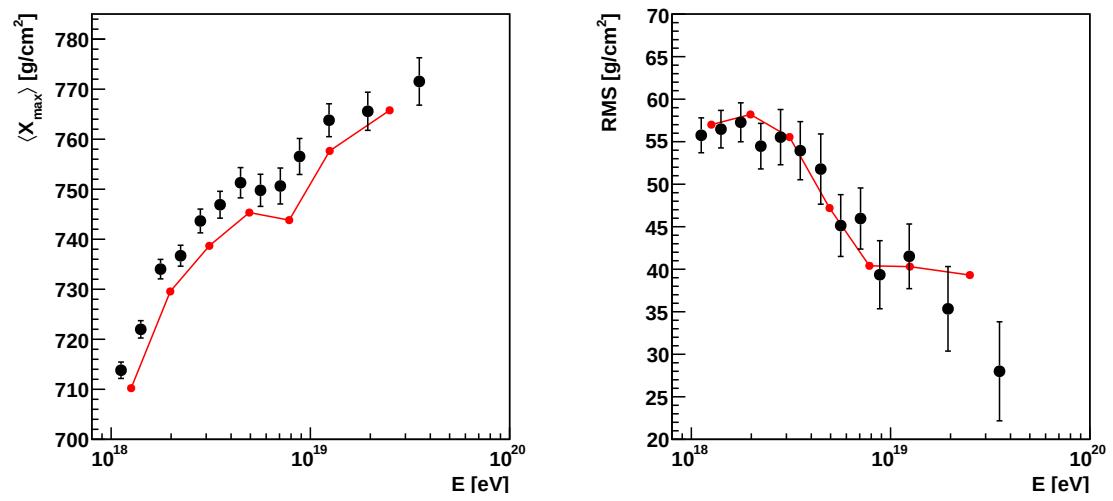

**Figure 1**: Comparing the Auger values for $\langle X_{max} \rangle$ (left) and RMS($X_{max}$) (right) [9] with the ones obtained from the 4-component model studied in this paper (red points). The 4-component model was obtained with a $\chi^2$ fit to the Auger $X_{max}$ distributions. There is 8 g/cm$^2$ difference between the fit and data in the $\langle X_{max} \rangle$ caused by low statistics in the tails of the $X_{max}$ distributions.

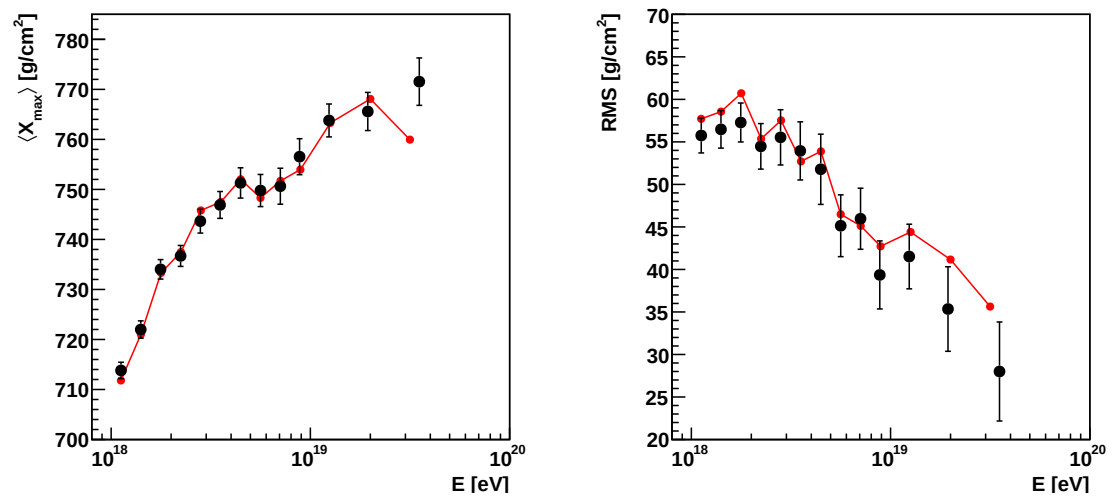

**Figure 2**: Same as Figure 1, but this time the 4-component model was obtained with a maximum likelihood fit to the Auger $X_{max}$ distributions.

Figure 4 compares the widths of the Auger data and the composition mixture used after the Telescope Array SD trigger bias. Again, the agreement is excellent over most energies. The bump in widths of the thrown composition mixture around $10^{18.3}$ - $10^{18.5}$ eV is driven by a deep tail of protons in the CORSIKA sample used to generate the Telescope Array shower library.

The Monte Carlo events were processed using Telescope Array hybrid reconstruction analysis software. Events are simulated and processed by the following procedure:

1. Showers are generated by CORSIKA and the SD trigger response is simulated.

2. The CORSIKA longitudinal shower profile for each shower is fitted to a Gaisser-Hillas function to determine the shower parameters.

3. A shower profile based upon the fitted shower parameters is generated and the TA fluorescence detector response including atmospheric, electronics, and geometrical acceptance is also simulated.

4. The shower geometry is fitted via the fluorescence profile and the shower-detector plane is measured.

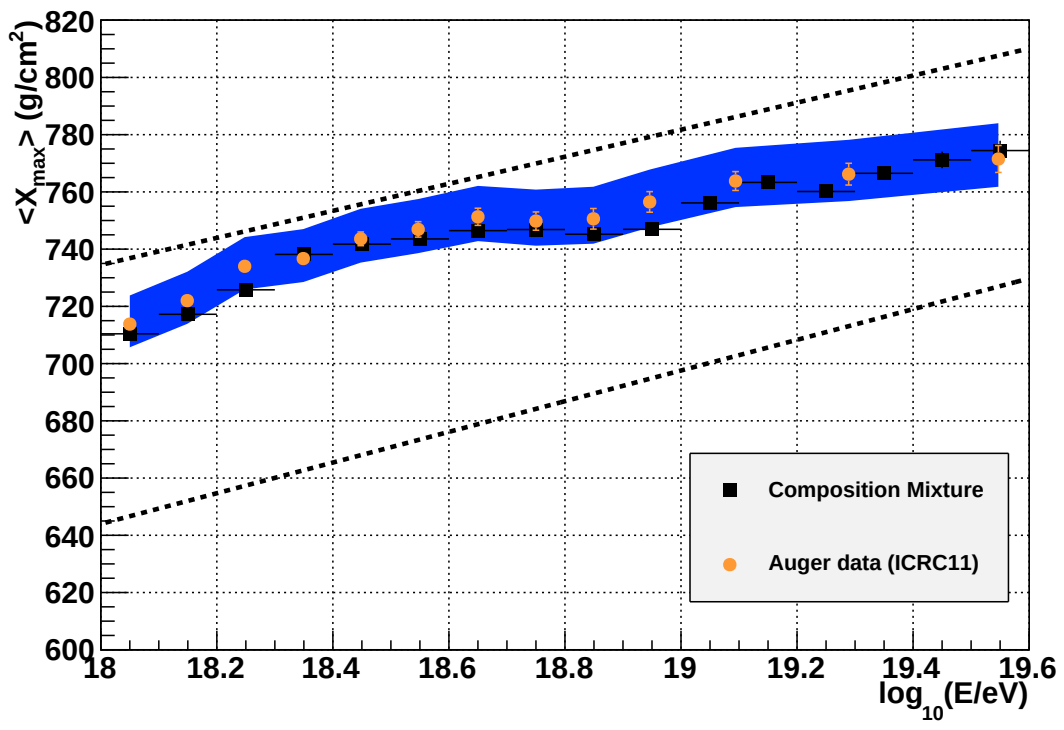

**Figure 3**: $\langle X_{\max}\rangle$ for the Auger composition mixture after Telescope Array SD trigger bias (black circles) compared to Auger data described in [9]. Dashed lines show QGSJetII proton and iron rails also from [9]. The blue band indicates Auger systematic uncertainties.

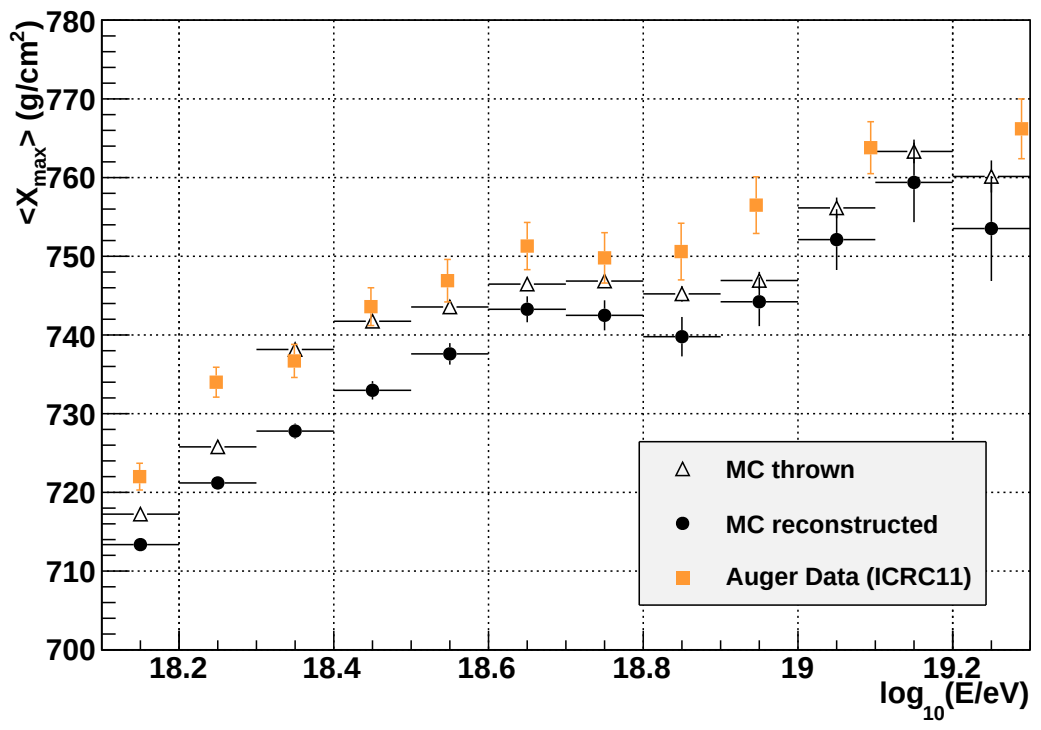

**Figure 5**: Reconstructed composition mixture $\langle X_{\max}\rangle$ compared with thrown $\langle X_{\max}\rangle$ after Telescope Array SD trigger bias and the most recent Auger composition data presented in [9].

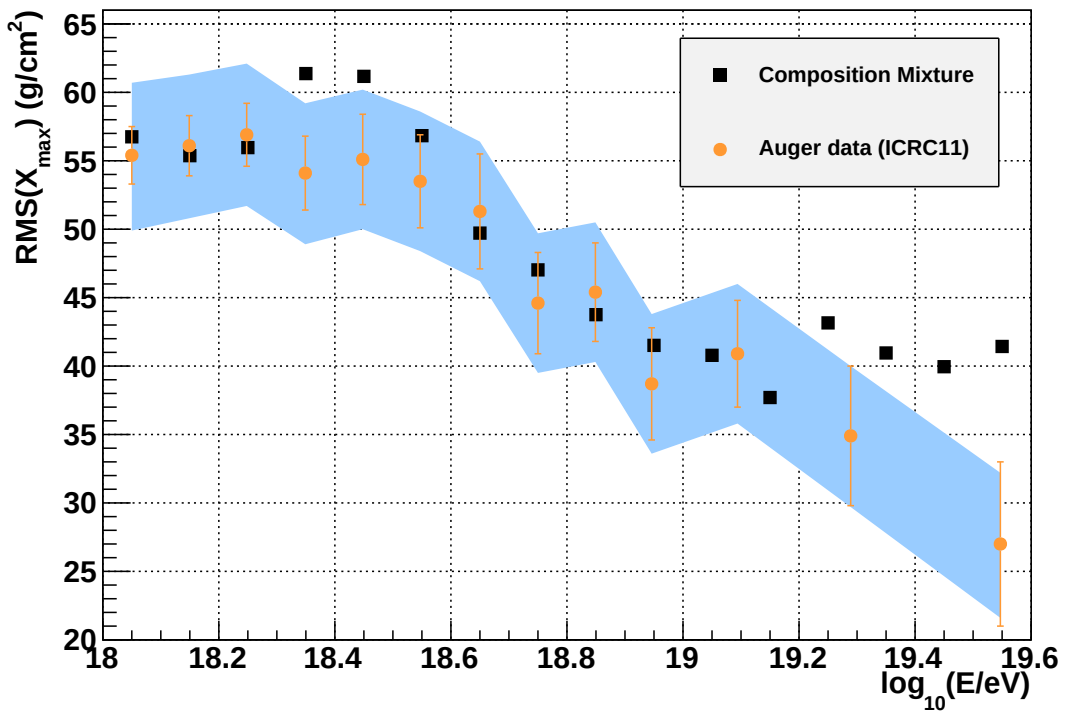

**Figure 4**: Thrown $X_{\max}$ RMS for the Auger composition mixture (black squares) including Telescope Array SD trigger bias compared to Auger data described in [9]. The blue band indicates Auger systematic uncertainties.

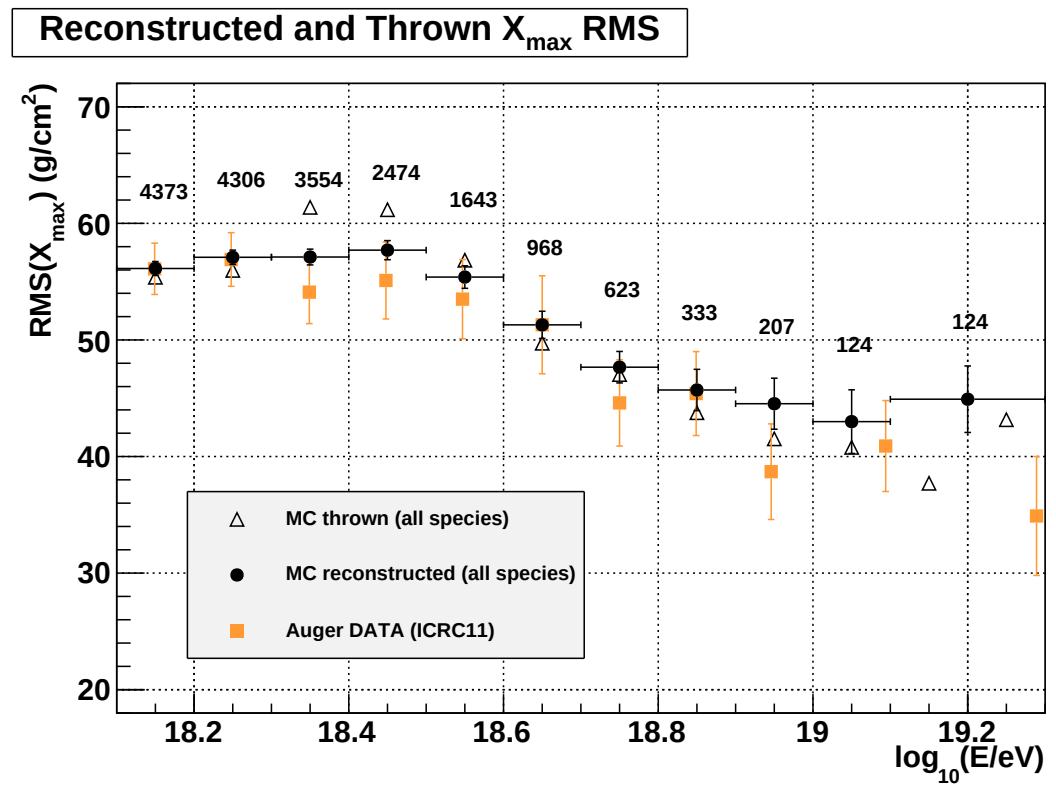

**Figure 6**: Reconstructed composition mixture $X_{\max}$ widths. The number or reconstructed events is also shown for each bin. Good agreement with the thrown distribution and the Auger data presented in [9] is seen here as well.

5. A fit to hybrid shower geometry is performed which combines the timing and geometric center of charge of the SD array, with the timing and geometry of the fluorescence detector that observed the event. This step is what makes the event a "hybrid event". If either the SD or FDs fail to trigger in an event, it can not be processed.

6. The shower profile is fitted via a reverse Monte Carlo method where the atmosphere, electronics, and geometrical acceptance of the shower are fully simulated.

The mean $X_{\max}$, after reconstruction of the composition mixture, is shown in figure 5. As has already been shown, the input distribution after SD trigger bias also agrees well with the Auger data. Comparison of the reconstructed widths (RMS) is shown in figure 6. Good agreement with the input distribution and with Auger data is also seen here. We have thus successfully reconstructed the expected features of an input spectrum composed of the given mixture: $\langle X_{\max}\rangle$ intermediate between protons and iron at the highest energies and widths that narrow as energy increases.

To see if TA hybrid reconstruction techniques can distinguish a spectrum composed purely of protons with one composed according to the Auger mix, a similar analysis was done using the spectrum composed purely of protons. The same spectral shape used for the mixed composition was also applied to the proton spectra and reconstructed using the same techniques. In the top panel of figure 7 $\langle X_{\max}\rangle$ for the reconstructed composition mixture is compared to $\langle X_{\max}\rangle$ for protons (iron reconstruction is included as reference). Over this energy range the mixture can be distinguished from protons and iron. A similar situation is shown in the bottom panel of figure 7 where the widths of the $X_{\max}$ distributions are compared. Above $10^{18.6}$ eV where the widths of the mixture $X_{\max}$ begin to narrow, no issues with Telescope Array reconstruction biases or acceptance preclude distinguishing a pure proton or pure iron spectrum from one that looks like the composition mixture.

Figure 8 shows the overall TA hybrid bias in $\langle X_{\max}\rangle$ for pure proton and for the Auger mix. The bias in both cases is nearly energy independent and it is found to be $11.5 \pm 0.9$ g/cm$^2$ for pure protons and $5.2 \pm 0.4$ g/cm$^2$ for the Auger mix. For this Figure, the total bias calculated for the Auger mix has the Telescope Array surface detector bias removed from it.

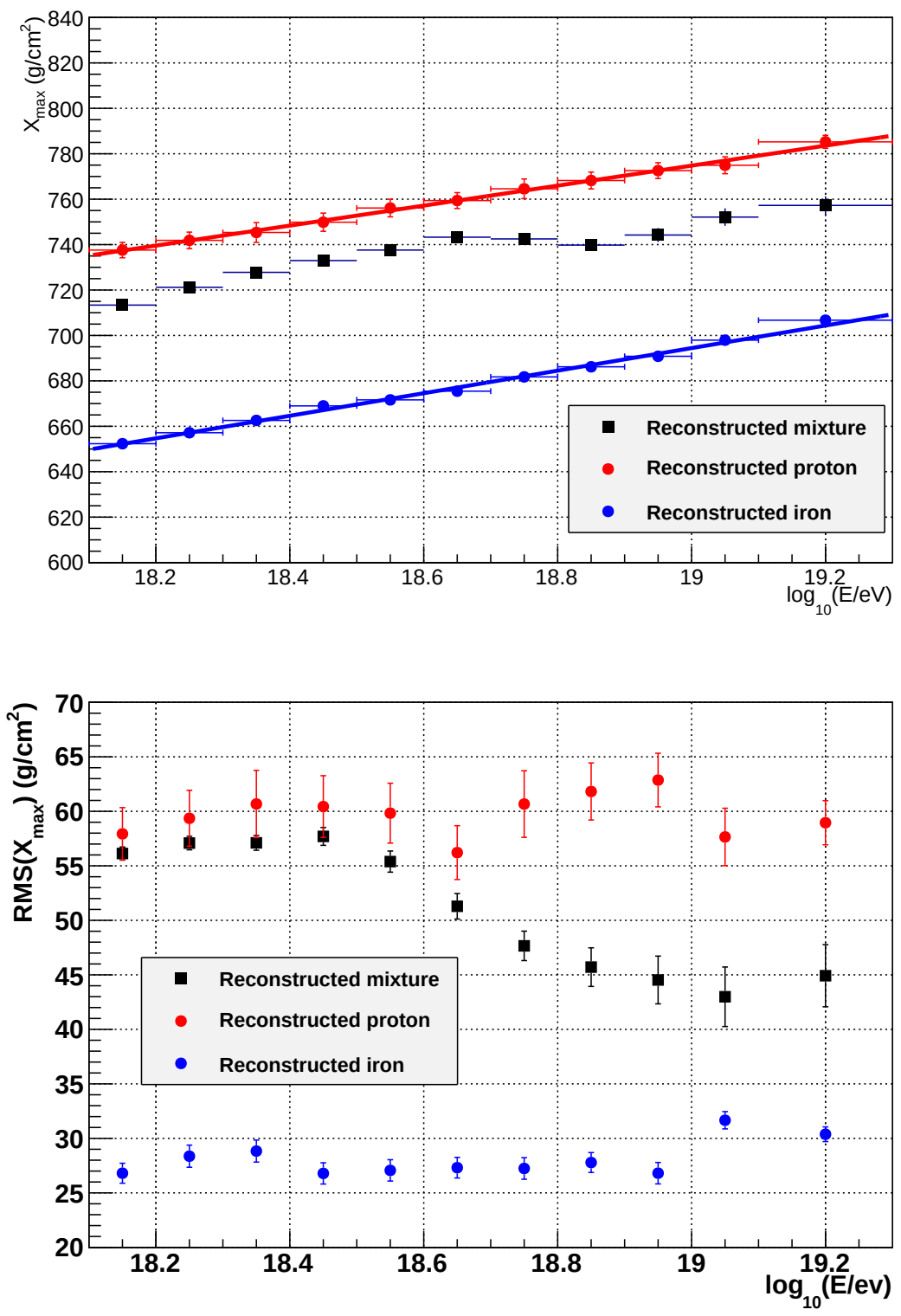

**Figure 7**: Comparison of $\langle X_{max} \rangle$ and widths of the reconstructed composition mixture and $\langle X_{max} \rangle$ and widths of pure proton and pure iron compositions. The three sets are clearly separated in each figure. The mixed composition can be distinguished from protons and iron using TA hybrid reconstruction. The fits to the reconstructed iron and proton $\langle X_{max} \rangle$ are also shown.

## 3 Conclusions

To begin to understand the apparent differences between Telescope Array/HiRes and Auger composition results, Auger has provided TA with an *ad hoc* model which fits Auger composition measurements. It consists of a 4-component mixture of protons, helium, nitrogen, and iron (the Auger mix) that varies with energy. Telescope Array generated a large Monte Carlo set based on the Auger mix, passed it through the full hybrid reconstruction analysis to obtain the expected $\langle X_{max} \rangle$ and $X_{max}$ widths for the Auger mix. In the same way the expectations for pure proton and pure iron were estimated. Figure 7 shows that the difference in the expected $\langle X_{max} \rangle$ between the Auger mix and the pure proton composition ranges from about 15 g/cm$^2$ at $10^{18.65}$ eV to about 30 g/cm$^2$ at $10^{18.85}$ eV.

The expected $\langle X_{max} \rangle$ and $X_{max}$ widths for the Auger mix and for the pure proton composition can be compared directly with the real Telescope Array hybrid results. Given the MC statistics generated in this simulation (e.g. 124 events in the energy bin of $10^{19}$), and the Telescope Array hybrid reconstruction biases and acceptances over $10^{18.1}$ - $10^{19.3}$ eV, Telescope Array can distinguish between the pure proton composition and the mixed composition provided by Auger (with at least 4 sigmas confidence level at $10^{19}$ eV). With adequate statistics in the data, Telescope Array will be able to distinguish between pure proton composition and the Auger mix composition.

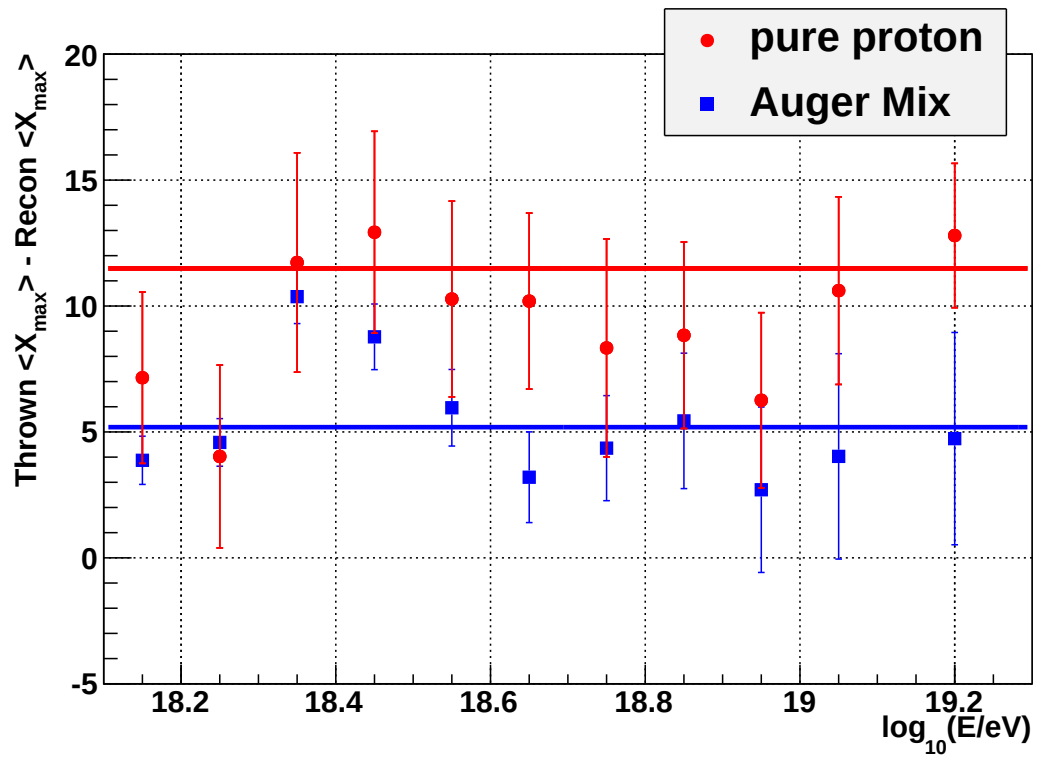

**Figure 8**: The difference between thrown $\langle X_{max} \rangle$ and reconstructed $\langle X_{max} \rangle$ for pure protons and the Auger mix. The total bias in $\langle X_{max} \rangle$ for the Auger mix is $5.2 \pm 0.4$ g/cm$^2$ (with TA SD trigger bias removed) and the total bias for pure protons is $11.5 \pm 0.9$ g/cm$^2$.

This Monte Carlo simulation assumes a given atmospheric model that could be slightly different from the one in real data. In this work we have not estimated the systematics due to uncertainties in the atmospheric model used in the Monte Carlo.

## 2 Acknowledgements

The successful installation, commissioning, and operation of the Pierre Auger Observatory would not have been possible without the strong commitment and effort from the technical and administrative staff in Malargüe. The Pierre Auger Collaboration is very grateful to the following agencies and organizations for financial support: Comisión Nacional de Energía Atómica, Fundación Antorchas, Gobierno De La Provincia de Mendoza, Municipalidad de Malargüe, NDM Holdings and Valle Las Leñas, in gratitude for their continuing cooperation over land access, Argentina; the Australian Research Council; Conselho Nacional de Desenvolvimento Científico e Tecnológico (CNPq), Financiadora de Estudos e Projetos (FINEP), Fundação de Amparo à Pesquisa do Estado de Rio de Janeiro (FAPERJ), São Paulo Research Foundation (FAPESP) Grants #2010/07359-6, #1999/05404-3, Ministério de Ciência e Tecnologia (MCT), Brazil; AVCR, MSMT-CR LG13007, 7AMB12AR013, MSM0021620859, and TACR TA01010517 , Czech Republic; Centre de Calcul IN2P3/CNRS, Centre National de la Recherche Scientifique (CNRS), Conseil Régional Ile-de-France, Département Physique Nucléaire et Corpusculaire (PNC-IN2P3/CNRS), Département Sciences de l'Univers (SDU-INSU/CNRS), France; Bundesministerium für Bildung und Forschung (BMBF), Deutsche Forschungsgemeinschaft (DFG), Finanzministerium Baden-Württemberg, Helmholtz-Gemeinschaft Deutscher Forschungszentren (HGF), Ministerium für Wissenschaft und Forschung, Nordrhein-Westfalen, Ministerium für Wissenschaft, Forschung und Kunst, Baden-Württemberg, Germany; Istituto Nazionale di Fisica Nucleare (INFN), Ministero dell'Istruzione, dell'Università e della Ricerca (MIUR), Gran Sasso Center for Astroparticle Physics (CFA), CETEMPS Center of Excellence, Italy; Consejo Nacional de Ciencia y Tecnología (CONACYT), Mexico; Ministerie van Onderwijs, Cultuur en Wetenschap, Nederlandse Organisatie voor Wetenschappelijk Onderzoek (NWO), Stichting voor Fundamenteel Onderzoek der Materie (FOM), Netherlands; Ministry of Science and Higher Education, Grant Nos. N N202 200239 and N N202 207238, The National Centre for Research and Development Grant No ERA-NET-ASPERA/02/11, Poland; Portuguese national funds and FEDER funds within COMPETE - Programa Operacional Factores de Competitividade through Fundação para a Ciência e a Tecnologia, Portugal; Romanian Authority for Scientific Research ANCS, CNDI-UEFISCDI partnership projects nr.20/2012 and nr.194/2012, project nr.1/ASPERA2/2012 ERA-NET, PN-II-RU-PD-2011-3-0145-17, and PN-II-RU-PD-2011-3-0062, Romania; Ministry for Higher Education, Science, and Technology, Slovenian Research Agency, Slovenia; Comunidad de Madrid, FEDER funds, Ministerio de Ciencia e Innovación and Consolider-Ingenio 2010 (CPAN), Xunta de Galicia, Spain; The Leverhulme Foundation, Science and Technology Facilities Council, United Kingdom; Department of Energy, Contract Nos. DE-AC02-07CH11359, DE-FR02-04ER41300, DE-FG02-99ER41107, National Science Foundation, Grant No. 0450696, The Grainger Foundation USA; NAFOSTED, Vietnam; Marie Curie-IRSES/EPLANET, European Particle Physics Latin American Network, European Union 7th Framework Program, Grant No. PIRSES-2009-GA-246806; and UNESCO.

The Telescope Array experiment is supported by the Japan Society for the Promotion of Science through Grants-in-Aids for Scientific Research on Specially Promoted Research (21000002) "Extreme Phenomena in the Universe Explored by Highest Energy Cosmic Rays" and for Scientific Research (19104006), and the Inter-University Research Program of the Institute for Cosmic Ray Research; by the U.S. National Science Foundation awards PHY-0307098, PHY-0601915, PHY-0649681, PHY-0703893, PHY-0758342, PHY-0848320, PHY-1069280, and PHY-1069286; by the National Research Foundation of Korea (2007-0093860, R32-10130, 2012R1A1A2008381, 2013004883); by the Russian Academy of Sciences, RFBR grants 11-02-01528a and 13-02-01311a (INR), IISN project No. 4.4509.10 and Belgian Science Policy under IUAP VII/37 (ULB). The foundations of Dr. Ezekiel R. and Edna Wattis Dumke, Willard L. Eccles and the George S. and Dolores Dore Eccles all helped with generous donations. The State of Utah supported the project through its Economic Development Board, and the University of Utah through the Office of the Vice President for Research. The experimental site became available through the cooperation of the Utah School and Institutional Trust Lands Administration (SITLA), U.S. Bureau of Land Management, and the U.S. Air Force. The Telescope Array Collaboration also wishes to thank the people and the officials of Millard County, Utah for their steadfast and warm support. The Telescope Array Collaboration gratefully acknowledges the contributions from the technical staffs of its home institutions. An allocation of computer time from the Center for High Performance Computing at the University of Utah is gratefully acknowledged.